\documentclass[12pt,preprint,xdvi]{emulateapj}

\usepackage{natbib}
\usepackage{graphicx}
\usepackage{times}
\usepackage{natbib}
\usepackage{graphicx}

\newcommand{\be}{\begin{equation}}
\newcommand{\ee}{\end{equation}}
\newcommand\eq{eq.}
\newcommand\eqs{eqs.}
\newcommand\fig{Fig.}
\newcommand\figs{Figs.}
\def\va{v_{\rm A}}
\def\bvec{{\bf B}}
\def\kvec{{\bf k}}
\def\vvec{{\bf v}}
\def\xvec{{\bf x}}
\def\rvec{{\bf r}}
\def\xhat{{\bf \hat{x}}}

\def\zhat{{\bf \hat{z}}}

\def\khat{{\bf \hat{k}}}
\def\half{\hbox{${1\over2}$}}

\def\zxpt{z_{\times}}

\def\oma{\omega_{\rm A}}

 \newcommand{\Frac}[2]{\frac{\raisebox{-0.5ex}{\ensuremath{#1}}}{#2}}

\newcommand{\op}[1]{\mathbb{#1}}

\begin{document}

\title{The Role of fast magnetosonic waves in the release and conversion via reconnection of energy stored by a current sheet}

\author{D.W. Longcope and L. Tarr}
\affil{Department of Physics, Montana State University,
  Bozeman, Montana 59717}



\begin{abstract}
Using a simple two-dimensional, zero-$\beta$ model, we explore the manner by which reconnection at a current sheet releases and dissipates free magnetic energy.  We find that only a small fraction (3\%--11\% depending on current sheet size) of the energy is stored close enough to the current sheet to be dissipated abruptly by the reconnection process.  The remaining energy, stored in the larger-scale field, is converted to kinetic energy in a fast magnetosonic disturbance propagating away from the reconnection site, carrying the initial current and generating reconnection-associated flows (inflow and outflow).  Some of this reflects from the lower boundary (the photosphere) and refracts back to the X-point reconnection site.  Most of this inward wave energy is reflected back again, and continues to bounce between X-point and photosphere until it is gradually dissipated, over many transits.  This phase of the energy dissipation process is thus global and lasts far longer than the initial purely local phase.  In the process a significant fraction of the energy (25\%--60\%) remains as undissipated fast magnetosonic waves propagating away from the reconnection site, primarily upward.  This flare-generated wave is initiated by unbalanced Lorentz forces in the reconnection-disrupted current sheet, rather than by dissipation-generated pressure, as some previous models have assumed.  Depending on the orientation of the initial current sheet the wave front is either a rarefaction, with backward directed flow, or a compression, with forward directed flow.
\end{abstract}


\section{Introduction}

Magnetic reconnection has been frequently proposed as the mechanism whereby magnetic energy, stored in the solar corona, is rapidly released in a solar flare.  In most current models, fast magnetic reconnection occurs at a current sheet where magnetic field lines of differing connectivity are brought into close enough proximity for a small-scale process, such as Ohmic diffusion or various kinetic effects, to forge new connections between them.  The current sheet, by carrying a net current, also stores the magnetic energy which the reconnection liberates.  This energy is not, however, co-located with the current sheet itself so the mechanism responsible for the reconnection electric field will not be the same one responsible for releasing or dissipating the magnetic energy.  Many investigations have been focussed on the former, local mechanism (flux transfer) while far fewer have addressed the latter, global mechanism (energy release).

\begin{figure}[htb]
\epsscale{0.8}
\centerline{(a)\plotone{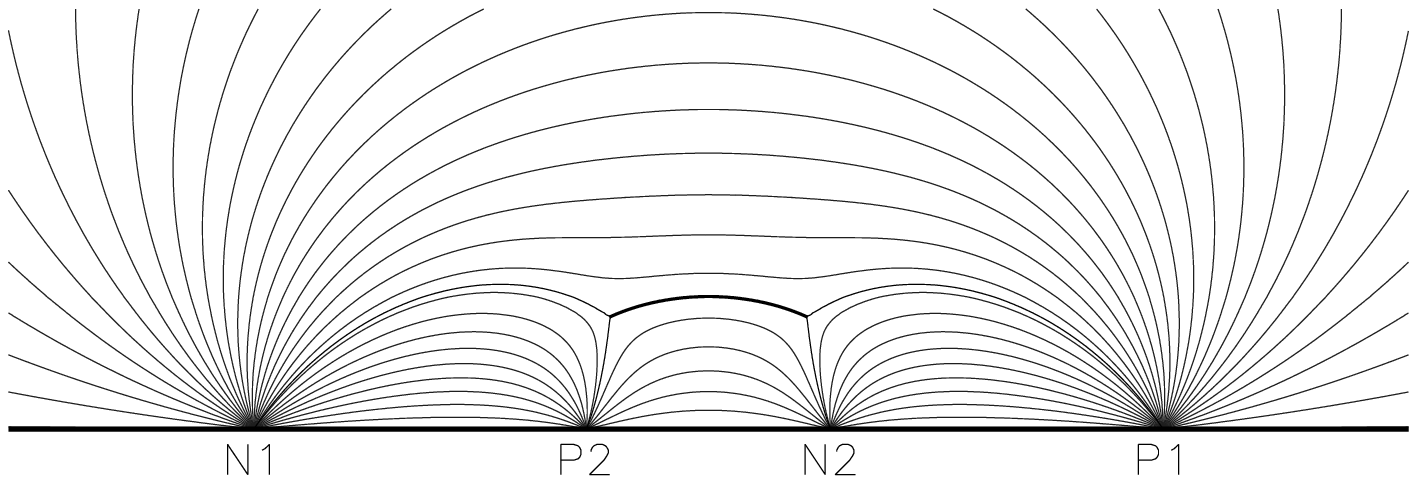}}
\centerline{(b)\plotone{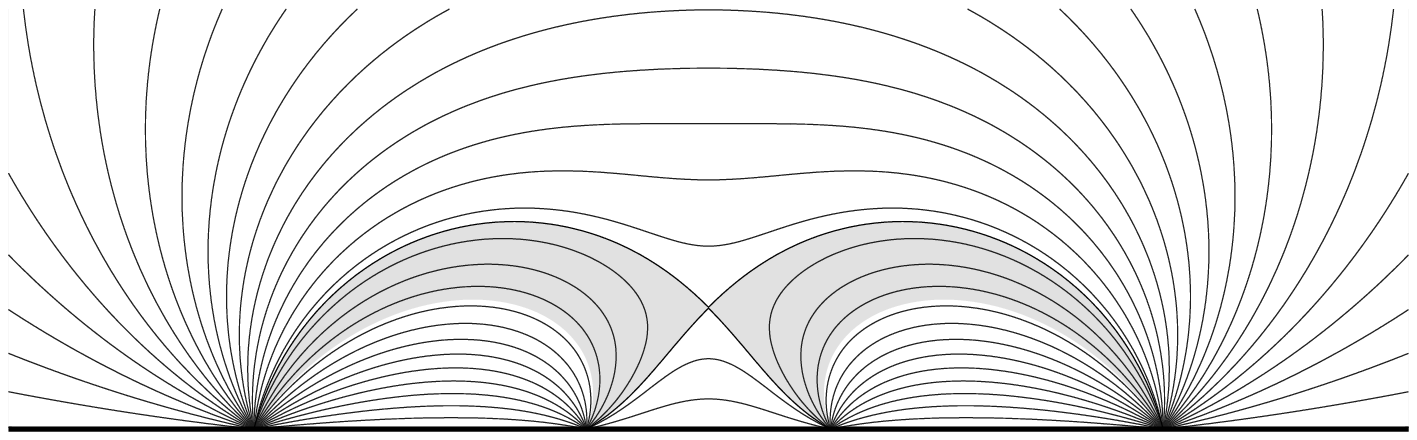}}
\centerline{(c)\plotone{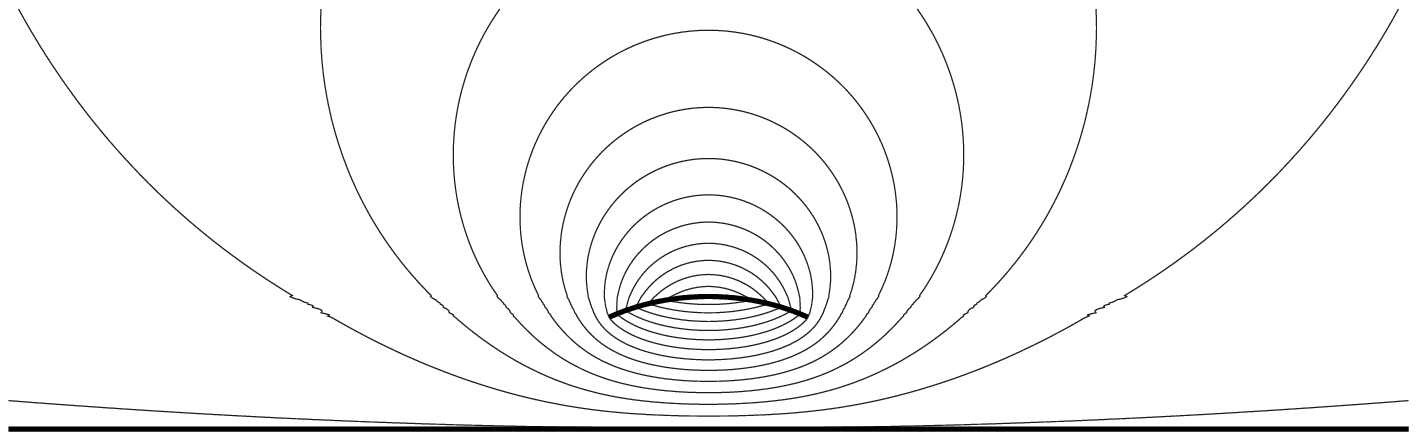}}
\caption{The field from four photospheric, magnetic sources.  (a) Equilibrium $\bvec(\xvec)$ state after emergence and incomplete reconnection of the bipole $P2$--$N2$.  The current sheet is shown as a bold arc connected to thinner lines: the separatrices.  (b) The potential field, $\bvec_0(\xvec)$, accessed by reconnection transfering flux $\Delta\psi$ into connection between $P2$--$N1$ and$P1$--$N2$; the new field lines fill the shaded regions.  (c)  The non-potential field, $\bvec-\bvec_0$, due to the current sheet.}
	\label{fig:qpl}
\end{figure}

A concrete illustration of the above issue is provided by the simple, two-dimensional quadupolar coronal field shown in \fig\ \ref{fig:qpl}.  A pair of sources, $P2$ and $N2$, have emerged underneath an older bipole, $P1$--$N1$.  If not all of the new flux is able to reconnect with the overlying flux, the state of minimum magnetic energy, $\bvec(\xvec)$, will contain a current sheet separating new from old flux as shown in \fig\ \ref{fig:qpl}a \citep[see][]{Heyvaerts1977,Priest2000}.  
The state of lowest possible energy is a potential field, $\bvec_0(\xvec)$, shown in \fig\ \ref{fig:qpl}b, in which additional flux $\Delta\psi$ (grey regions) interconnects the new and old polarities.  A current sheet, carrying current $I_{\rm cs}$, exists to maintain a connectivity different from the potential field, as in \fig\ \ref{fig:qpl}a.  For this reason the net current $I_{\rm cs}\sim\Delta\psi$ \citep{Longcope2001b,Longcope2004}.  The current sheet will be reduced, or eliminated entirely, as flux is transferred across the sheet into domain connecting $P1$--$N2$ and $P2$--$N1$ (the shaded regions).  It is in this way the local topological changes at the current sheet may have energetic consequences.

The free magnetic energy of the topologically constrained magnetic field is computed by subtracting the energy of the unconstrained (i.e.\ potential) field,
\begin{eqnarray}
  \Delta E_M &=& {1\over8\pi}\int|\bvec|^2\, d^3x ~-~{1\over8\pi}\int|\bvec_0|^2\, d^3x \nonumber \\
  &=& {1\over8\pi}\int|\bvec-\bvec_0|^2\, d^3x ~~,
\end{eqnarray}
where the integrals are over the entire coronal volume ($z>0$).\footnote{A cross term involving the integral of
$(\bvec-\bvec_0)\cdot\bvec_0$ can be seen to vanish after expressing $\bvec_0=-\nabla\chi$ and integrating by parts.}  The final expression shows that the free energy is equivalent to the energy of the {\em non-potential} component, 
$\bvec(\xvec)-\bvec_0(\xvec)$, due to the current sheet with homogeneous conditions at the lower boundary
($B_z-B_{0z}=0$).  This energy will naturally decrease as the current $I_{\rm cs}$ is decreased by reconnection.  The non-potential field, shown in \fig\ \ref{fig:qpl}c, extends far beyond the sheet itself.  At large distances it decreases inversely with distance similar to the field from a simple wire.

It is evident from the form of the non-potential field (\fig\ \ref{fig:qpl}c) that changes in the current sheet must propagate great distances in order to release the stored magnetic energy.  Much of the volume over which the free magnetic energy is stored is not in magnetic contact with the current sheet, so neither Alfv\'en waves nor slow magnetosonic waves (nor shocks) can change the field directly.   Yet if the current in the sheet is to change, that change must be reflected in the distant field, in accordance with Amp\`ere's law.  How this occurs is unlikely to depend too critically on the mechanism responsible for the flux transfer which must, according to all current theories, occur on small scales.  Thus it should be possible to study the mechanism of energy release using a arbitrary form of rapid flux transfer at the current sheet.

Such an analysis was performed by \citet{Longcope2007e} using a current sheet situated at a two-dimensional X-point, in a zero-$\beta$ plasma, whose reconnection was effected by a sudden increase in Ohmic resistivity.  They found that the disruption of the current sheet launched a cylindrical current shell at the leading edge of a fast magnetosonic wave (FMW).  The shell contained almost all of the current formerly carried by the current sheet.   It therefore left in its wake (i.e.\ inside the shell) a nearly potential magnetic field.  The non-potential magnetic energy of the initial field was converted to kinetic energy of the wave's flow.  Remarkably, the Ohmic dissipation responsible for initiating the wave, and thus releasing the stored magnetic energy, directly dissipated a rather small fraction of the stored energy.  This puzzling result, that large resistivity does not result in large Ohmic losses, can be anticipated from the arguments above: even after it diffuses, the current sheet occupies a very small volume and therefore has access to very little magnetic energy.

The flow field in the FMW has a quadrupolar structure familiar in steady reconnection models: inflows along one axis (the vertical axis in \fig\ \ref{fig:qpl}) and outflows along the other (horizontal).  Previous studies of transient magnetic reconnection models have found fast magnetosonic rarefaction waves to be the drivers of inflow \citep{Lin1994,Heyn1996,Nitta2001}.  These analyses, set on infinitely long current sheets, focussed on the inner reconnection region rather than the front of the wave.  Since \citet{Longcope2007e} considered a sheet carrying finite current they were able to analyze its full global propagation and energy release.  In particular, their model revealed that the kinetic energy of the inflow must be supplied by decreasing the free magnetic energy stored in the initial equilibrium.  In their unbounded domain this kinetic energy is not subsequently converted to heat.

The finite current sheet studied by \citet{Longcope2007e} was situated in an unbounded domain, which therefore contained an infinite amount of free magnetic energy.  The FMW would propagate indefinitely, converting this stored energy to kinetic energy at a uniform rate.  The wave energy could therefore become arbitrarily large in comparison to  the energy directly dissipated at the current sheet.  

In order to make contact with previous work by \citet{Craig1991} and \citet{Hassam1992}, \citet{Longcope2007e} briefly considered the effect of a boundary: a concentric cylindrical conductor.  This reflected (perfectly) the outward-propagating FMW back inward.  The inward wave collapsed on the X-point, as described by \citet{Craig1991} and \citet{Hassam1992} permitting still more Ohmic dissipation to occur there.  The majority of the wave's energy was, however, {\rm reflected} once more by the X-point.  The wave was therefore trapped between a perfectly reflecting outer conductor and an imperfectly reflecting X-point.  The only losses were at the X-point, so eventually all energy was in fact dissipated at the X-point through Ohmic dissipation.  The characteristic time for dissipation depended on the round-trip transit time between reflectors.  Owing to the exponential increase of the Afv\'en speed with distance, this transit time scales logarithmically with the dissipation scale, and thus logarithmically with the resistivity.  The ability of an X-point to Ohmically dissipate magnetic energy in logarithmic time was the most significant result of the studies \citet{Craig1991} and \citet{Hassam1992}, and appears to suggest that Ohmic dissipation alone is capable of all magnetic energy dissipation in a flare.

The upshot of this reasoning is that the initial disruption of the current sheet by Ohmic diffusion will dissipate a
small fraction of its magnetic energy, but that repeated reflections can result in the complete dissipation of all stored energy  at the X-point.  To achieve the latter end, however, previous authors have assumed a conducting boundary completely surrounding the X-point.  In a more realistic geometry, such as that of \fig\ \ref{fig:qpl}, there is a conducting boundary at the photosphere ($z=0$) but it does not completely surround the X-point.  

\citet{McLaughlin2006} studied the linearized, $\beta=0$ dynamics of a quadrupolar magnetic field with a planar lower boundary like \fig\ \ref{fig:qpl}.  Rather than initiate a wave by reconnection, they launched a FMW by {\em fiat} from the lower boundary.  They found that a fraction of the wave (they report 40\%) was refracted into the null, while the remainder continued to propagate away.  This suggests that the photospheric lower boundary may indeed be less effective at mediating energy dissipation than are the concentric cylindrical conductors, although it is not clear that the same fraction would apply to a wave launched outward from the null rather than from the lower boundary.  

\citet{McLaughlin2006} omitted diffusion from their model but assumed its effect would be to dissipate all the wave energy refracted into the null.    \citet{Craig1991} and \citet{Hassam1992} found to the contrary that the main effect of resistivity is to {\em reflect } the wave back outward from the null.  It was only through repeated reflection that the wave energy could be ultimately dissipated.  If only 40\% were reflected back at each step (as suggested by \citet{McLaughlin2006}), then the final dissipation would be far less than 40\% initially directed toward the null.

The lack of 100\% reflection back to the null point appears to pose a difficulty for the system reaching a potential state.  It was found by \citet{Longcope2007e} that flux transfer continues at the X-point long after the majority of the current sheet had disappeared.  As a result $\Delta\psi$ continues to decrease, {\em below zero}, bringing the system {\em away} from the potential field state ($\Delta\psi=0$).   When the wave reflected from the cylindrical boundary, however, it reversed this flux transfer, bringing $\Delta\psi$ back upward.  While it continued to overshoot the potential value, numerous reflections led to its gradual convergence to zero.  If the initial wave is only partially reflected back to the null point, as the results of \citet{McLaughlin2006} suggest, then it is unclear how $\Delta\psi$ would ever converge to zero, and the system achieve a potential field.

To more fully understand energy release and dissipation we seek to determine what fraction of the FMW launched from the null point will be reflected back to the X-point in this more realistic configuration.  We must also know how much of the partially reflected wave will ultimately dissipate at the X-point and what fraction will reflect once more.

This analysis, performed below, shows that the lowest frequency components of the wave are reflected almost completely by the photospheric boundary.  This results in the current at the X-point being entirely eliminated, and $\Delta\psi\to0$, after numerous reflections.  It also results in the direct dissipation of a significant fraction of the free energy (between 40\% and 75\% in one example we consider).  This does not occur immediately through the local reconnection mechanism, Ohmic diffusion in our case, but rather it requires repeated reflections from the photosphere and thus takes many transit times to achieve.  The remainder of the free energy (25\% -- 60\%) is emitted by the reconnection as FMW at higher frequencies.  These waves form a number of pulses propagating primarily vertically upward, away from the photosphere.  The flow in these wave forms is either a rarefaction or a compression depending on the orientation of the initial current sheet (horizontal or vertical, respectively).  Such fast magnetosonic disturbances have long been known to accompany flares, but until now no quantitative reconnection model has predicted what fraction of the free magnetic energy they accounted for.

We present the model calculation for a two-dimensional, quadrupolar field, like the one in \fig\ \ref{fig:qpl}.  In the next section we specify the geometry of the model field, and quantify the free magnetic energy stored in advance of reconnection.  In the following section we analyze the dynamical behavior in the vicinity of the X-point, and describe how this is matched to the external field, including the photospheric boundary.  Section \ref{sec:num} presents numerical solutions to the external response, including the emission of a FMW upward.  The numerical solutions cannot be continued for the many FMW-transit times of the full solution.  Instead we use them to characterize the reflection from the photosphere and then use the reflection coefficient to synthesize a solution for long times in \S \ref{sec:long_time}.  This long time solution is used to quantify the ultimate fate of the free energy.

\section{Model of reconnection in a quadrupolar field}

\subsection{The quadrupolar equilibrium}

We begin with a potential magnetic field created by four photospheric sources, as depicted in \fig\ \ref{fig:qpl}b.  The field is expressed in terms of a flux function $\bvec(y,z) = \nabla A\times\xhat$, whose potential version is
\be
  A_0(y,z) ~=~ \sum_j{\psi_j\over\pi} \tan^{-1}\left( {z+d\over y - y_j} \right) ~~,
  	\label{eq:A0}
\ee
where source $j$ is located at $(y_j,-d)$ and has flux (per ignorable length) $\psi_j$.  Singular line sources are located a distance $d$ below the photospheric boundary ($z=0$) so that the field is non-singular within the model ($z>0$).  The sum runs over positive sources $j=1,2$ and negative sources $j=-1,-2$.  We create reflectional symmetry about the $z$ axis by taking $\psi_{-1}=-\psi_1$ and $\psi_{-2}=-\psi_2$ as well as $y_{-1}=-y_1$ and $y_{-2}=-y_2$.  
Except for scaling, the quadrupole is described by the two dimensionless 
parameters $\psi_1/\psi_2$ and $y_2/y_1$; we take the latter to fall in the range $(-1,0)$ so that $j=-2$ and $+2$ create the inner bipole whose orientation is opposite to the outer bipole.

The potential field generated by (\ref{eq:A0}) will vanish at a single coronal X-point located at $(0,\zxpt)$, where 
\be
  \zxpt ~=~ -d ~+~ y_1\,\sqrt{{|y_2|\over y_1}\,{(\psi_2y_1+\psi_1y_2)\over(\psi_1y_1+\psi_2y_2)}} ~~,
  	\label{eq:zxpt}
\ee
provided this is a real positive number.  There are two conditions for the expression to be real.  First, that the denominator inside the radical must remain positive, $|y_2|/y_1 < \psi_1/\psi_2$, which is equivalent to requiring the overlying field to be directed opposite to the lower bipole.  The second is that the numerator be positive, $|y_2|/y_1 < \psi_2/\psi_1$, in order that the the field direction not reverse along the line between $P2$ and $N2$.  (When this is violated two null points are located symmetrically along the line between $P2$ and $N2$ --- at $z=-d$.)  Once these conditions are met it is still necessary that $\zxpt>0$ in order that the null point is actually above the lower boundary at $z=0$.

In the immediate vicinity of the null point the potential field flux function, \eq\ (\ref{eq:A0}), is approximated by its Taylor expansion
\be
  A_0(y,z) ~\simeq~ A_0(0,\zxpt) ~+~ \half B'_0\, [\,y^2 - (z-\zxpt)^2\, ] ~~,
  	\label{eq:A0_X}
\ee
where the magnetic shear at the X-point is
\be
  B'_0 ~=~\left.{\partial^2A_0\over\partial y^2}\right\vert_{(0,\zxpt)} ~=~-\sum_j{2\psi_j y_j(\zxpt+d)\over 
  \pi[(\zxpt+d)^2+y_j^2]^2} ~~.
  	\label{eq:Bp0}
\ee
This is the same structure analyzed by 
\citet{Longcope2007e}, and we will use those results over an {\em inner region}, extending to a small distance $r_i$ from the null point.   Within this region we assume the expansion (\ref{eq:A0_X}) to be a good approximation to the full field, \eq\ (\ref{eq:A0}).

We add to the potential field, $\bvec_0$, the non-potential field from a current sheet carrying current 
\be
  I_{\rm cs} ~=~{1\over 4\pi}\oint\bvec\cdot d{\bf l} ~~.
\ee
The full field, $\bvec(\xvec)$, is a force-free equilibrium with a singular current, similar to that in \fig\ \ref{fig:qpl}a.  The current will be taken small enough that the sheet is a straight line of half-length $\Delta=2\sqrt{|I_{\rm cs}|/B'_0}$, centered at $(0,\zxpt)$.  If $I_{\rm cs}>0$ the sheet will be horizontal, similar to \fig\ \ref{fig:qpl}a; if $I_{\rm cs}<0$ it will be vertical.  We will assume the current is small enough that 
$\Delta\ll r_i$. 

In order to preserve the distribution of photospheric flux an image current sheet, with opposite sign, will be added at $z=-\zxpt$.  The combination will yield a flux function value on the current sheet \citep{Longcope2001b}
\begin{eqnarray}
  A_{\rm cs} &\simeq& A_0(0,\zxpt) ~+~ I_{\rm cs}\ln( 16e\zxpt^2/\Delta^2 ) \nonumber \\
  &=& A_0(0,\zxpt) ~+~ I_{\rm cs}\ln( 4\zxpt^2eB'_0/|I_{\rm cs}| ) ~~.
\end{eqnarray}
where $e$ is the base of the natural logarithm.
The result of the current sheet is thus to introduce an additional flux 
$\Delta\psi=I_{\rm cs}\ln( 4\zxpt^2eB'_0/|I_{\rm cs}|)$ connecting $P2$ to $N2$.  The smallness of the current sheet means that outside the inner region radius ($r_i$) the full field may be approximated
\begin{eqnarray}
  A(y,z) &\simeq& A_0(y,z) ~-~ I_{\rm cs}\ln\left[ {y^2 + (z-\zxpt)^2\over y^2 + (z+\zxpt)^2} \right] \nonumber \\[7pt]
  &\simeq& A_0(y,z) ~+~ 2I_{\rm cs}\ln(2\zxpt/r)~~,
  	\label{eq:A_out}
\end{eqnarray}
where $r$ is the distance from $(0,\zxpt)$.

The free energy of the magnetic field can be computed by integrating the electromagnetic work required to ramp the current up to its final value \citep{Longcope2001b}
\begin{eqnarray}
  \Delta E_M &=& \int\limits_0^{\Delta\psi}I\,d(\Delta\psi) ~=~ \half I_{\rm cs}^2\ln( 4\zxpt^2 B'_0e^{1/2}/|I_{\rm cs}| ) 
  \nonumber \\
  &=& \half I_{\rm cs}^2\ln( 16\zxpt^2 e^{1/2}/\Delta^2)  ~~.
  	\label{eq:DW_tot}
\end{eqnarray}
The amount of this free energy outside the inner region is
\begin{eqnarray}
  \Delta E^{\rm (out)}_M &=&
  {1\over8\pi}\int_{r>r_i}|\nabla A-\nabla A_0|^2\, d^2x \nonumber \\
  &\simeq& \half I_{\rm cs}^2\ln( 4\zxpt^2/r_i^2) ~~.
	\label{eq:DW_out}
\end{eqnarray}
It is due to the lower boundary, and the image current, that these expressions are finite, while those of \citet{Longcope2007e} were infinite.

\subsection{Linear dynamics}

Having assumed a small current sheet we can consider its contribution to be a small perturbation, $A_1$, to the potential quadrupolar field \citep[see][for extensive discussion of this approximation]{Longcope2007e}.  This evolves, along with a linear velocity field, $\vvec_1$, according to the momentum and resistive induction equations of resistive MHD
\begin{eqnarray}
  {\partial \vvec_1\over\partial t} &=& -{\nabla A_0\over4\pi\rho_0}\,\nabla^2 A_1 ~~,
  	\label{eq:momentum} \\
  {\partial A_1\over\partial t} &=& -\vvec_1\cdot\nabla A_0 ~+~\eta\nabla^2 A_1 ~~,
  	\label{eq:induction}
\end{eqnarray}
where $\eta$ is the Ohmic resistivity and $\rho_0$ the plasma's initial mass density.  The pressure force has been dropped under the assumption that plasma $\beta$ is very low.

Following \citet{Longcope2007e}, we introduce an inductive electric field variable, 
$U_1=\vvec_1\cdot\nabla A_0=-\xhat\cdot(\vvec_1\times\bvec_0)$, to obtain the pair of scalar equations
\begin{eqnarray}
  {\partial U_1\over\partial t} &=& -\va^2(\xvec)\,\nabla^2 A_1 ~~, \label{eq:U1} \\
  {\partial A_1\over\partial t} &=& - U_1 ~+~\eta\nabla^2 A_1 ~~,
  	\label{eq:A1}
\end{eqnarray}
where $\va=|\nabla A_0|/\sqrt{4\pi\rho_0}$ is the Alfv\'en speed in the potential field.  For simplicity we henceforth take the initial mass density, $\rho_0$, to be uniform.  The system is solved beginning with $A_1(y,z,0)$ generating the current sheet.  The resistivity is then ``turned on'' to some fixed value at $t=0$.  The current sheet then diffuses initiating the dynamical evolution studied by \citet{Longcope2007e}.

Within the inner region the potential magnetic field increases linearly with distance from its X-point,
\be
  |\nabla A_0| ~=~ B'_0\, r ~~~,~~~r ~=~\sqrt{ y^2 + (z-\zxpt)^2} ~<~ r_i ~~.
  	\label{eq:B-xpt}
\ee
This means the Alfv\'en speed increases linearly as well, $\va=\oma\,r$, where 
$\oma=B_0'/\sqrt{4\pi\rho_0}$ is a frequency characteristic of the null point.  This frequency defines a resistive length scale $\ell_{\eta}=\sqrt{\eta/\oma}$, inside which diffusion is a dominant effect.  We propose that the enhanced resistivity turned on at $t=0$ be large enough that $\ell_{\eta}\gg\Delta$.  This will assure that the current sheet is entirely disrupted by the diffusive process \citep{Longcope2007e}.  This is the way we model simply a mode of fast magnetic reconnection in order to focus on the external response.

While the diffusive length is much larger than the current sheet, we take it to be small compared to the potential field X-point.  This permits us to define the inner region such that $r_i\gg\ell_{\eta}$, and the resistive term may be dropped from \eq\ (\ref{eq:A1}) outside of it ($r>r_i$).  We will then be able to match the inner solution of \citet{Longcope2007e} to the non-diffusive solution of the outer region.  We will first consider, in the next section, solutions in the interface region $r\simeq r_i$ at which the matching must occur.  We then produce the complete solution of the inner region in a form suitable for matching.

\section{Dynamics of the inner region}

\subsection{The interface region}

In the vicinity of the interface, $r=r_i$, we may neglect the resistive term ($r_i\gg\ell_{\eta}$) and use the X-point expansion of the Alfv\'en speed.  We take advantage of the approximate cylindrical symmetry to expand
\begin{eqnarray}
  A_1(r,\phi,t) &=& \sum_m \hat{A}^{(m)}(r,t)\,e^{i m\phi} ~~,\\
  U_1(r,\phi,t) &=& \sum_m \hat{U}^{(m)}(r,t)\,e^{i m\phi} ~~,
\end{eqnarray}
where the sum extends over both positive and negative values of $m$.  The modal coefficients then satisfy the equations
\begin{eqnarray}
    {\partial \hat{U}^{(m)}\over\partial t} &=& -\oma^2\left(r{\partial\over\partial r}\right)^2\hat{A}^{(m)} 
    + m^2\oma^2\, \hat{A}^{(m)} ~~,\\
  	\label{eq:Um}
  {\partial \hat{A}^{(m)}\over\partial t} &=& -\hat{U}^{(m)} ~~,
  	\label{eq:Am}
\end{eqnarray}
The axisymmetric equations ($m=0$) are satisfied by arbitrary wave forms 
\be
  \hat{U}^{(0)} \pm \oma\,r{\partial\hat{A}^{(0)}\over\partial r} ~=~ W_{\pm}( \oma t\mp\ln r) ~~,
  	\label{eq:W_def}
\ee
propagating outward ($W_+$) or inward ($W_-$).  Because the wave speed increases linearly with radius, a particular phase point in the wave accelerates exponentially as it travels: $r\sim e^{\pm \oma t}$.  

Higher mode numbers ($|m|\ge1$) satisfy a Klein-Gordon equation in $\ln r$, with a cut-off frequency,  
$|m|\oma$.   Frequencies below the cut-off are evanescent at the interface.  We demonstrate below that higher-$m$ perturbations of low frequency incident on the X-point will be reflected from the interface.

The initial magnetic field in the interface region is given by the second term of \eq\ (\ref{eq:A_out}), for which 
$r\partial A_1/\partial r=-2I_{\rm cs}$.   This is an axisymmetric field in equilibrium ($U_1=0$) consisting of counter-propagating waves of the form
\be
  W_-(x) ~=~ -W_+(x) ~=~ 2\oma I_{\rm cs} ~~.
\ee
It seems that in the absence of resistivity the X-point and current sheet function as a perfect reflector, at least at zero frequency.  It reflects an incoming wave with a sign flip in order to maintain the current-sheet equilibrium.  We see below that the onset of resistive diffusion changes the nature of this reflection, thereby eliminating the current sheet.

Integrating the Poynting flux around the entire the interface at $r_i$ yields the power carried outward by the waves
\be
  P_i ~=~ {1\over 2}\sum_m \left.[\hat{U}^{(m)}]^*r{\partial \hat{A}^{(m)}\over\partial r}\right\vert_{r=r_i} ~~.
\ee
The axisymmetric contribution to this power
\be
  P^{(0)}_i ~=~ {1\over 8\oma}\Bigl[\, W_+^2 \,-\, W_-^2\, \Bigr] ~~,
  	\label{eq:P0}
\ee
is the difference in energy carried in and out across the interface.

\subsection{The inner region: current disruption}

\begin{figure}[htb]
\epsscale{1.2}
\centerline{\plotone{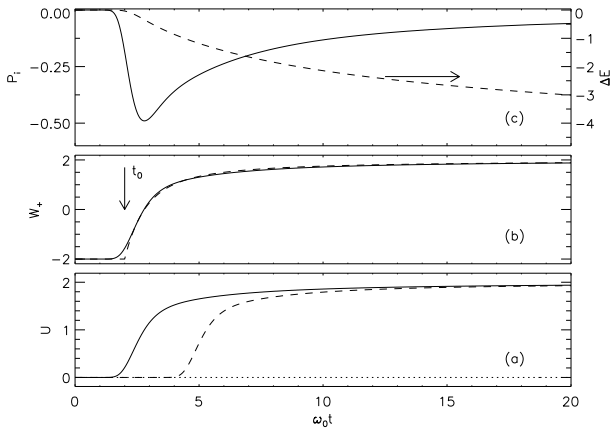}}
\caption{External response to current sheet dissipation after \citet{Longcope2007e}.  (a) The axisymmetric electric field, 
$\hat{U}^{(0)}$, scaled to $\oma I_{\rm cs}$,  at $r=12\ell_{\eta}$ (solid) and $r=150\ell_{\eta}$ (dashed) for comparison.  (b) The outgoing wave, $W_+$ at  $r=12\ell_{\eta}$ (solid), scaled to $\oma I_{\rm cs}$,  
and the fit (dashed) from \eq\ (\ref{eq:W-LP07}) using the value of $t_0$ shown by the arrow.  (This is the only free parameter in the fit.)  (c) The axisymmetric contribution to the Poynting flux, scaled to $\oma I_{\rm cs}^2$, at  $r=12\ell_{\eta}$ (solid).  Its time-integral (dashed), the total energy transferred, is scaled to $I_{\rm cs}^2$ and plotted against the axis on the right.}
	\label{fig:LP07}
\end{figure}

\citet{Longcope2007e} performed a thorough analysis of the dynamics following the diffusive disruption of the current sheet; pertinent results are summarized in \fig\ \ref{fig:LP07}.   The external effect of the current disruption is an axisymmetric FMW propagating outward.  The axisymmetric electric field, $\hat{U}^{(0)}$, 
at a fixed position (\fig\ \ref{fig:LP07}a) rises from zero as the FMW reaches it.  They found the form of the wave 
to be\footnote{There is a typographical error in the text following \eq\ (30) in \citet{Longcope2007e}.  It should have read $r_0=\sqrt{2}\ell_{\eta}$.  Equation (\ref{eq:W-LP07}) corrects that error.}
\be
  W_+(x) ~\simeq~ \left\{\begin{array}{lcl} -2\oma I_{\rm cs} &~~,~~& x<\oma t_0 \\[7pt]
  2\oma I_{\rm cs}\left[1-
  \displaystyle{1 - e^{-2(x-\oma t_0)} \over x-\oma t_0}\right]
  &~~,~~& x\ge \oma t_0 \end{array}\right.
  	\label{eq:W-LP07}
\ee
where $t_0$ is the time required for the wave to be initiated and reach the nominal position $r=1$.  This semi-empirical fit, plotted as a dashed curve over the actual wave form in \fig\ \ref{fig:LP07}b, is a reasonably good match.

Their analysis assumed an infinite exterior domain from which a steady incoming wave, $W_-=2\oma I_{\rm cs}$, persisted indefinitely.  The outgoing wave in \eq\ (\ref{eq:W-LP07}), switches from being negative, opposite to the incoming wave, to positive, $W_+\to 2\oma I_{\rm cs}$, matching the incoming wave (see \fig\ \ref{fig:LP07}b).  This represents a change from a static magnetic field with no velocity or electric field (an equilibrium with $U_1=0$) to a steady electric field, $U_1=2\oma I_{\rm cs}$ (see \fig\ \ref{fig:LP07}a), and vanishing magnetic field perturbation (i.e.\ the potential field).  Evidently the change in resistivity (i.e.\ its turn-on) produces this change in the X-point reflection properties.

The Poynting flux becomes negative as the FMW passes a given radius (see \fig\ \ref{fig:LP07}c).  It can be seen from 
\eq\ (\ref{eq:P0}) that it reaches a peak value, 
$P_i=-\oma I_{\rm cs}^2/2$, at the instant $W_+=0$.  After that $W_+\to W_-$ causing the Poynting flux to decrease asymptotically as $t^{-1}$.  The same inward Poynting pulse is observed, with some travel delay, at all distances from the current sheet independent of the value of resistivity responsible for reconnection.  

There is a final aspect of this wave solution worthy of note.  As the solution approaches the steady electric field, 
$U_1\to2\oma I_{\rm cs}$, \eq\ (\ref{eq:Am}) implies the flux function will become uniform, but with a steadily decreasing value
\be
  {\partial A_1\over\partial t} ~\simeq~-2\oma I_{\rm cs} ~~.
\ee
This steady flux transfer, dictated by the external solution, is accommodated by a persistent current density confined to an ever-shrinking neighborhood of the X-point well inside the diffusive radius $\ell_{\eta}$.  The central current density remains fixed, so the total current decreases as $t^{-1}$, the square of the shrinking diameter.  This current, amidst a steady electric field $U\sim 2\oma I_{\rm cs}$, produces the asymptotic Poynting flux seen in \fig\ \ref{fig:LP07}c.  The FMW imposes an inflow and the residual internal solution represents its accommodation at small scales.

\subsection{The inner region: response to outer dynamics}

The complete dynamics of the inner region, including resistive diffusion, is best treated after Fourier transforming
\be
  \hat{A}^{(m)}_{\omega}(r) ~=~ \int_{-\infty}^{\infty}\hat{A}^{(m)}(r,t)\,e^{i\omega t}\, dt ~~.
\ee
The Fourier coefficient must then satisfy the equation
\begin{eqnarray}
  \left[ \, \left(r{\partial\over\partial r}\right)^2 - m^2 \,\right]
  \hat{A}^{(m)}_{\omega} &-& {i\omega \eta\over r^2}\left[ \, \left(r{\partial\over\partial r}\right)^2 - m^2 \,\right]
  \hat{A}^{(m)}_{\omega}
  \nonumber \\
   ~+~ {\omega^2\over\oma^2} \hat{A}^{(m)}_{\omega} &=& 0 ~~.
\end{eqnarray}
\citet{Hassam1992} found an analytic solution for the case $m=0$ which was regular as $r\to0$.  This solution can be generalized to arbitrary $m$,
\be
  \hat{A}^{(m)}_{\omega}(r) ~=~ r^m\,F\left( a, a^*; m+1;\,-i{\oma r^2\over\omega\ell_{\eta}^2} \right) ~~,
\ee
where $F$ is the hypergeometric function \citep{Abramowitz1964} and
\be
  a ~=~ {1\over 2}\left[ m  + i \sqrt{ (\omega/\oma)^2 - m^2 } \right] ~=~ \half[ m + i\varpi_m ]~~.
\ee

For frequencies above cut-off, $\omega^2>m^2\oma^2$, the factor $\varpi_m$ is real and 
the hypergeometric function can be expanded well outside the diffusion region, $r\gg\ell_{\eta}$,
\begin{eqnarray}
  \hat{A}^{(m)}_{\omega}(r) &\simeq& \left({i\oma^2\over\eta\omega}\right)^{-m/2}\,\Biggl[
  X_m\,\left( i{\oma r^2\over\omega\ell_{\eta}^2}\right)^{i\varpi_m/2} \nonumber \\
  &+&  ~~X^*_m\,\left( i{\oma r^2\over\omega\ell_{\eta}^2}\right)^{-i\varpi_m/2}\,\Biggr] ~~,
	\label{eq:A_inner}
\end{eqnarray}
where $X_m=m!\Gamma(i\varpi_m)/a\Gamma^2(a)$.  The factors
\begin{eqnarray}
  &~& \left( i{\oma r^2\over\omega\ell_{\eta}^2}\right)^{\pm i\varpi_m/2} ~=~ \nonumber \\
  &=& \exp\Bigl[ \, \pm i\varpi_m\ln(r/\ell_{\eta}) \,
  \pm\, \half i \varpi_m\ln(i\oma/\omega) \, \Bigr] ~~.
	\label{eq:Xfactors}
\end{eqnarray}
can be identified with outward (upper sign) and inward (lower sign) propagating components, assuming that $\varpi_m$ is real and positive.  An analytical continuation through complex frequencies from one branch of the expression onto the other would require an exchange between what have been here designated as ``inward'' and ``outward'' components, thereby inverting the expression for the reflection coefficient defined below.   

The ratio of the terms corresponding to outward and inward waves in \eq\ (\ref{eq:A_inner}) gives a complex reflection coefficient
\begin{eqnarray}
  \hat{R}^{(m)}(\omega) &=& -{X_m\over X_m^*}\,\exp[ i\varpi_m\ln(i\oma/\omega)] \nonumber \\
  &=&
  -{X_m\over X_m^*}\,e^{-i\varpi_m\ln(\omega/\oma)}\,\exp[ -\hbox{${\pi\over 2}$}\varpi_m ] ~~,
  	\label{eq:Rm}
\end{eqnarray}
where a minus sign is introduced to represent the reflection of $U_1$ rather than $A_1$.  The magnitude of the coefficient represents the ratio of amplitudes between the incoming and the reflected wave,
\begin{eqnarray}
  |\hat{R}^{(m)}(\omega)| &=& \exp[ -\hbox{${\pi\over 2}$}\varpi_m ]  \nonumber \\
  &=& \exp\Bigl[ -\hbox{${\pi\over 2}$}\sqrt{(\omega/\oma)^2- m^2} \Bigr] ~~.
  	\label{eq:R_mag}
\end{eqnarray}
Resistivity at the X-point damps all waves above the cut-off frequency: $|\hat{R}|<1$.  At frequencies approaching the cut-off the X-point becomes a more perfect reflector; at frequencies below the cut-off it is a perfect reflector.  

The current sheet provides an initial condition $A_1(r,\phi,0)$ containing all even modes, $m=0,\pm2,\pm4,\dots$ 
\citep[and appendix]{Longcope2007e}.  The disruption of the current sheet occurs at frequencies near and below the characteristic frequency, $\oma$.  For all values $|m|\ge2$ this evolution will be below the cut-off frequency so virtually no power will emerge from the inner region in any mode except the axisymmetric mode, $m=0$.  This means, as we demonstrate more precisely in an appendix, that the free energy in non-axisymmetric modes is directly dissipated by resistivity, and the axisymmetric energy alone leaves the X-point region.  The axisymmetric mode will reflect off outer boundaries, i.e.\ the photosphere, to produce new waves propagating back into the X-point.  These waves will not be axisymmetric, but will contain power in frequencies similar to those in the initial wave, \eq\ (\ref{eq:W-LP07}).  The non-axisymmetric components will therefore reflect from the X-point.  We will henceforth focus on the axisymmetric, $m=0$, in the inner region.

The complex phase in the axisymmetric version expression of (\ref{eq:Rm}) represents the phase difference between the outward and inward waves, referred to position $r=\ell_{\eta}$.  Evaluation of that phase shows it to be nearly zero (i.e.\ $\hat{R}^{(0)}\simeq e^{-\pi\omega/2\oma}$) suggesting the waves reflect off the diffusive region.  In order to provide coupling between inner and outer regions we introduce an axisymmetric reflection coefficient referred to radius $r_i$
\be
  \hat{R}_i (\omega)~=~ \hat{R}^{(0)}(\omega)\,\exp[\,2i(\omega/\oma)\ln(r_i/\ell_{\eta})\,] ~~.
  	\label{eq:Ri}
\ee
This will have the same magnitude, \eq\ (\ref{eq:R_mag}), but a phase representing the round-trip transit from $r_i$ to the diffusive region.

An inward-propagating axisymmetric wave, with Fourier transform $\hat{W}_-(\omega)$, will produce an outward wave, 
$\hat{W}_+=\hat{R}_i\hat{W}_-$.  Using this in expression (\ref{eq:P0}) we find the total energy crossing the interface outward,
\begin{eqnarray}
  \Delta E &=& \int\limits_{-\infty}^{\infty}P_i\, dt ~=~
  {1\over 8\oma}\int\limits_{-\infty}^{\infty}( W_+^2 - W_-^2) \, dt\nonumber \\
   &=& -{1\over 16\pi}\int\limits_{-\infty}^{\infty}(1
  -|\hat{R}_i|^2)|\hat{W}_-|^2\, {d\omega\over\oma} ~~.
\end{eqnarray}
This negative value is the net energy dissipated at the X-point.  Using \eq\ (\ref{eq:R_mag}) in this expression
shows frequencies $\omega>\oma/\pi$ to be damped most effectively at the X-point.

\subsection{Coupling to the outer solution}

The outer region, while not described by a simple X-point field, is nevertheless described  by linear dynamics which can be summarized by a reflection coefficient, $\hat{R}_o$,  found in next section.  An outward propagating wave, 
$\hat{W}_+(\omega)$, will give rise to a reflected inward wave $\hat{W}_-(\omega)$; both waves are evaluated at $r=r_i$.  The outer reflection coefficient is then defined by
\be
  \hat{R}_o(\omega) ~=~ {\hat{W}_-(\omega)\over\hat{W}_+(\omega)} ~~.
  	\label{eq:Ro}
\ee
By direct analogy with $\hat{R}_i$, we may associate $1-|\hat{R}_o|^2$ with the energy lost to the outer region, although here it will be by a failure to reflect (i.e.\ free radiation) rather than Ohmic dissipation.

Coupling the inner and outer regions at the interface $r_i$ produces a kind of cavity between the two reflectors.  Requiring the waves in both expressions to match (i.e.\ the inward-wave from outside is equal to the outward wave from the X-point) gives the resonance condition
\be
  \hat{R}_o(\omega)\hat{R}_i(\omega) ~=~ 1 ~~.
  	\label{eq:res}
\ee
This will be satisfied by a set of frequencies, $\omega_n$, constituting the resonant frequencies of the cavity.  We have shown above that $|\hat{R}_i(\omega)|<1$ for real, non-zero, frequencies; the external domain will satisfy the condition $|\hat{R}_o(\omega)|\le1$ for real frequencies.  The resonance condition, \eq\ (\ref{eq:res}) will thus be satisfied only for complex frequencies.

An example of this cavity analysis is provided by the analyses of \citet{Craig1991} and \citet{Hassam1992}.  Their external potential field was a perfect X-point, just like the inner field.  (In this simple case they did not need to separate the field into inner and outer regions.)  The field was bounded by a cylindrical boundary at $r=L$ which was a perfect conductor: $\hat{U}^{(0)}(L)=0$.  This generates a perfect reflection
\be
  \hat{R}_o(\omega) ~=~ \exp[\,2i(\omega/\oma)\ln(L/r_i) - i\,\pi\, ] ~~,
\ee
where the $-i\pi$ accounts for the inversion of $\hat{U}$ at the conductor.  Taking the phase of $\hat{R}^{(0)}$ to be zero, as it approximately is, gives a resonance condition for the cylindrical cavity
\begin{eqnarray}
  \hat{R}_o(\omega)\hat{R}_i(\omega) &=& \exp\Bigl\{\,2i(\omega/\oma)[ \ln(L/\ell_{\eta})
  + i\pi/4 ]  - i\, \pi \, \Bigr\} \nonumber \\
  &=& 1 ~~,
\end{eqnarray}
for ${\rm Re}(\omega)>0$.\footnote{This restriction follows from the designation of inward and outward waves, as explained in the text following \eq\ (\ref{eq:Xfactors}).}  The resonant frequencies are thus
\begin{eqnarray}
  \omega_n &=& {\oma\,\pi(n+\half)\over \ln(L/\ell_{\eta}) + i\pi/4} \nonumber \\
  &\simeq&
  {\oma\,\pi(n+\half)\over \ln(L/\ell_{\eta})} ~-~i {\oma\,\pi^2(n+\half)\over 4\ln^2(L/\ell_{\eta})} ~~,
\end{eqnarray}
for $n=0,\,1,\,\dots$.  The damping time of each mode is therefore proportional to the square of the logarithm of the
Lundquist number, $S=L^2\oma/\eta=L^2/\ell_{\eta}^2$, as originally found by \citet{Hassam1992}.  The real part of the resonant frequency is set by the requirement that an odd number of quarter waves fit between the node at $r=L$ and the anti-node at $r=\ell_{\eta}$, separated by transit time $\ln(L/\ell_{\eta})/\oma$.

The quality of the cavity
\be
  Q ~=~ -{{\rm Re}(\omega_n)\over{\rm Im}(\omega_n)} ~\simeq~ {4\ln(L/\ell_{\eta})\over\pi} ~=~
  {2\ln S\over \pi} ~~,
\ee
is roughly the number of times the wave must reflect off the X-point before it is damped.  Lundquist numbers of astrophysical scale, $S\sim10^{12}$, would require a relatively large number of reflections off the perfectly conducting cylinder before the X-point would be able to dissipate all the energy Ohmically.

\section{The Dynamics of the external field}
\label{sec:num}

\subsection{Numerical solution}

We produce a numerical solution to the time-dependent equations of linear dynamics, \eqs\ (\ref{eq:U1}) and (\ref{eq:A1}), in the external region.  Diffusive effects are limited to the inner region so we set $\eta=0$ in the external equations.  The field $A_1$ and $U_1$ are represented on a uniform $L_y\times L_z$ rectilinear grid.  Both fields are located on the same points, and the Alfv\'en speed from the potential field, $\va(\xvec)$, is computed on these points as well.  The Laplacian of $A_1$ is computed using a simple finite-difference operation which is second-order accurate on the uniform grid.  Along the outer boundaries, $y=\pm L_y/2$, $z=0$ and $z=L_z$, the conducting boundary condition $U_1=0$ is imposed.  The grid is made large enough, compared to $y_1$ and $\zxpt$, that reflections from the upper and lateral boundaries do not return the X-point during a solution.  The reflection from the $z=0$ boundary is the focus of the investigation.

A propagating wave condition is applied at an interior boundary, $r\simeq r_i$.  This is done by identifying two contiguous, closed paths of grid points called $C_-$ and $C_+$.  Each path is approximately circular, at radius $r_-$ and $r_+$, such that $(r_++r_-)/2\simeq r_i$.  The paths are nested, meaning no extraneous grid points lie between them.  Averages are computed along each path as a means of identifying the axisymmetric components $\hat{U}^{(0)}$ and 
$\hat{A}^{(0)}$ at both radii,
\be
  \bar{U}_{\pm} ~=~{1\over N_{\pm}}\sum_{C_{\pm}}U_1 ~~~~,~~~~
  \bar{A}_{\pm} ~=~{1\over N_{\pm}}\sum_{C_{\pm} }A_1 ~~~~,
\ee
where $N_{\pm}$ is the number of grid points along path $C_{\pm}$.
At each time step the values of $\bar{A}_-$, $\bar{A}_+$ and $\bar{U}_+$ are computed from the present solution.  These  are used to set new values for $U_1$ along the inner path, $C_-$.  This is done using a finite-difference version of \eq\ (\ref{eq:W_def}),
\be
  \bar{U}_- = -\bar{U}_+ - \oma(r_++r_-){\bar{A}_+-\bar{A}_-\over r_+-r_-} + 2\,W_+(\oma t - \ln r_i) ~~,
\ee
where $W_+$ is the prescribed wave-form being fed into the solution.  At each point along the inner path, $C_-$, the value of $U_1$ is set to the value $\bar{U}_-$ computed for that step.  The value of $A_1$ along both paths is advanced using \eq\ (\ref{eq:A1}), with $\eta=0$, which requires only values of $U_1$ at the same point --- no information from neighboring points.  The solution on points inside the inner path is never used and is therefore irrelevant (the equations are solved there anyway since time-stepping is performed over the entire rectilinear grid).

The procedure above imposes a prescribed, axisymmetric wave, $W_+$, propagating away from the inner region and into the external solution.  It also serves as a perfect absorber of axisymmetric waves incident on the inner region.  The form of the incident wave passing out of the computational domain can be computed after the fact using the same finite-difference version of  \eq\ (\ref{eq:W_def})
\be
  W_-(\oma t-\ln r_i) ~=~ \half( \bar{U}_++ \bar{U}_- ) ~-~ \half \oma(r_++r_-){\bar{A}_+-\bar{A}_-\over r_+-r_-} ~~.
  	\label{eq:W_inc}
\ee
Since $U_1$ is set to the same value at all points along $C_-$, any wave component with $m\ne0$ is perfectly reflected from the inner boundary.  This is an approximation to the actual coupling, but as shown above it is a reasonably good one for low frequencies.  

The value of $\eta$ appears nowhere in the equations of the outer region, and the solution is thus applicable to any value provided $\ell_{\eta}\ll r_i$.  Its only effect on the outbound waveform, \eq\ (\ref{eq:W-LP07}), is in the delay time $t_0$.  We eliminate this final dependance by setting
\be
  W_+(\oma t -\ln r_i) ~=~ 2\oma I_{\rm cs}\left[1-
  {1 - e^{-2\oma t} \over \oma t}\right] ~~~,~~~ t>0 ~~,
  	\label{eq:W_in}
\ee
so the wave enters the external domain at the beginning of the solution, $t=0$.  When applying the numerical solution to a given $\eta$, the current-sheet disruption will have occurred (i.e.\ $\eta$ will have been switched on) at the negative time
$t=-\ln(r_i^2\oma/\eta)/2\oma$.

The initial conditions are found by solving Laplace's equation, $\nabla^2A_1=0$ numerically subject to the homogeneous outer boundary conditions and holding $A_1$ at a constant value along the inner path $C_-$.  This solution is then rescaled so that
\be
  \half (r_1+r_0){\bar{A}_1-\bar{A}_0\over r_1-r_0} ~=~ -2 ~~,
\ee
which gives a positive current of unit magnitude.  The amplitude of the linear solution is irrelevant, so in this way we normalize it to $I_{\rm cs}$.

\begin{figure}[htb]
\epsscale{1.2}
\centerline{\plotone{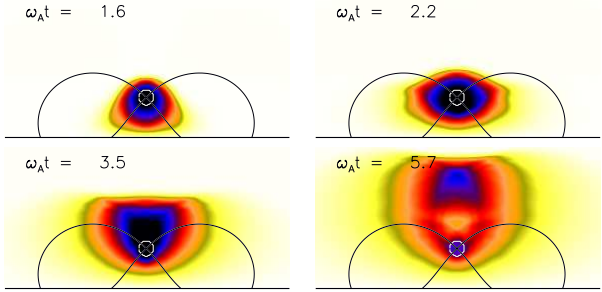}}
\caption{The solution to linear dynamics amidst a quadrupolar potential field.  $U_1(y,z)$ at $\oma t=1.6$, 2.2, 3.5 and 5.7, from upper left (color online).  The white curve is the inner boundary $C_-$.  Gray curves are the separatrices of the potential field.}
	\label{fig:Usol}
\end{figure}

Figure \ref{fig:Usol} shows a typical solution to the linear equations.  This case uses a quadrupolar field with 
$\psi_2/\psi_1=4/5$ and $y_2/y_1=-1/2$.  The source depth is set to $d=0.1y_1$, and \eqs\ (\ref{eq:zxpt}) and (\ref{eq:Bp0}) give $\zxpt=0.4\, y_1$ and $B'_0=1.92\,\psi_1/y_1^2$.  We use a $701\times300$ point grid with 
conducting boundaries at $z=0$, $z=3\,y_1=7.5\,\zxpt$ and $y=\pm 2.5\,y_1$.  (The grid spacing is therefore $\Delta y=\Delta z=0.01\,y_1$.)  The inner boundary is taken to be $r_i=0.07y_1$; the inner path $C_-$ is a white curve in the figure.  The four panels show $U_1(y,z)$ at four successive times in the solution, depicting the initial departure of the wave (upper left), its reflection from the photosphere (upper right), and then encountering the X-point upon reflection (lower panels).  The central region is positive in the initial phases.

\begin{figure}[htb]
\epsscale{1.2}
\centerline{\plottwo{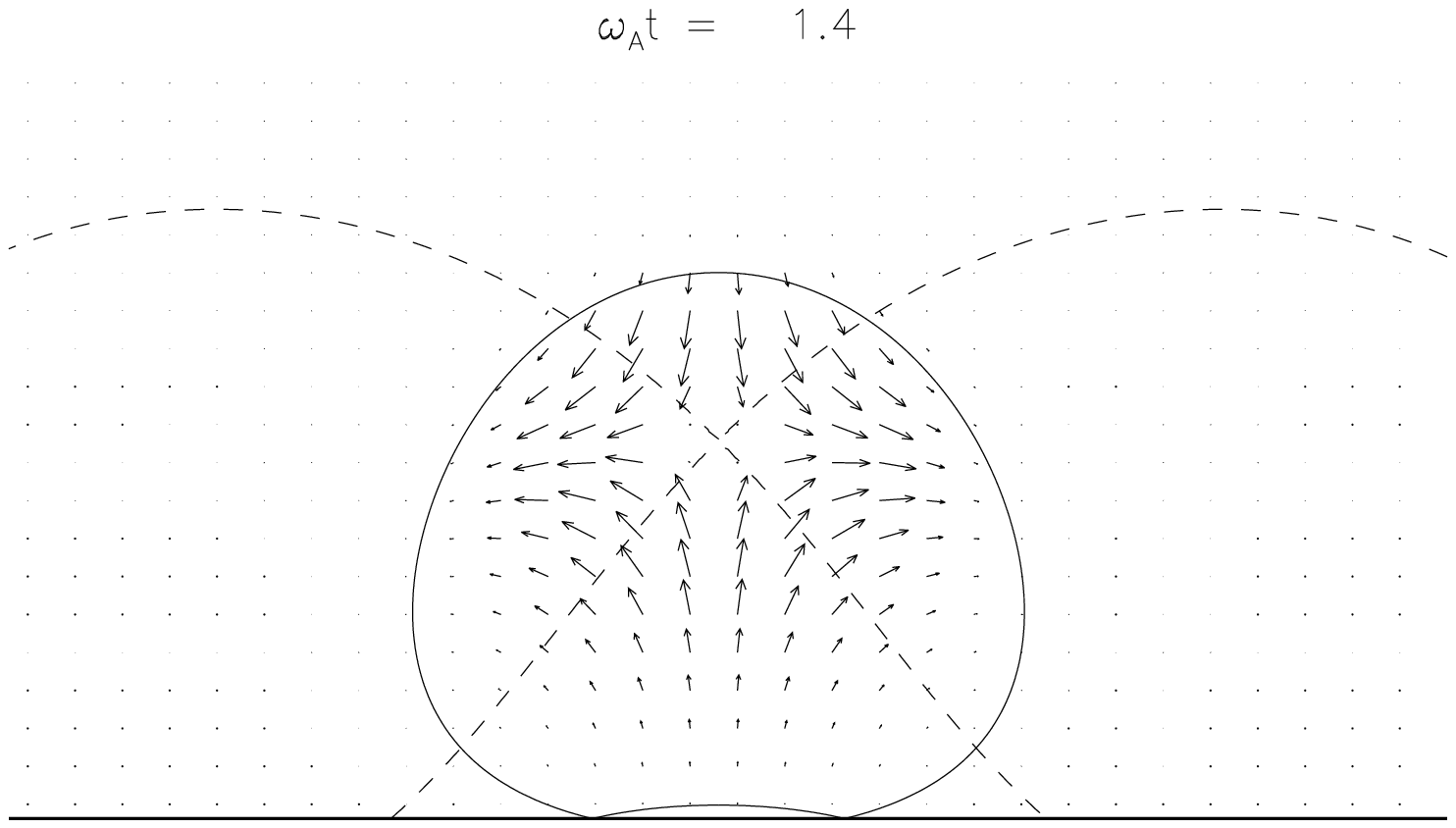}{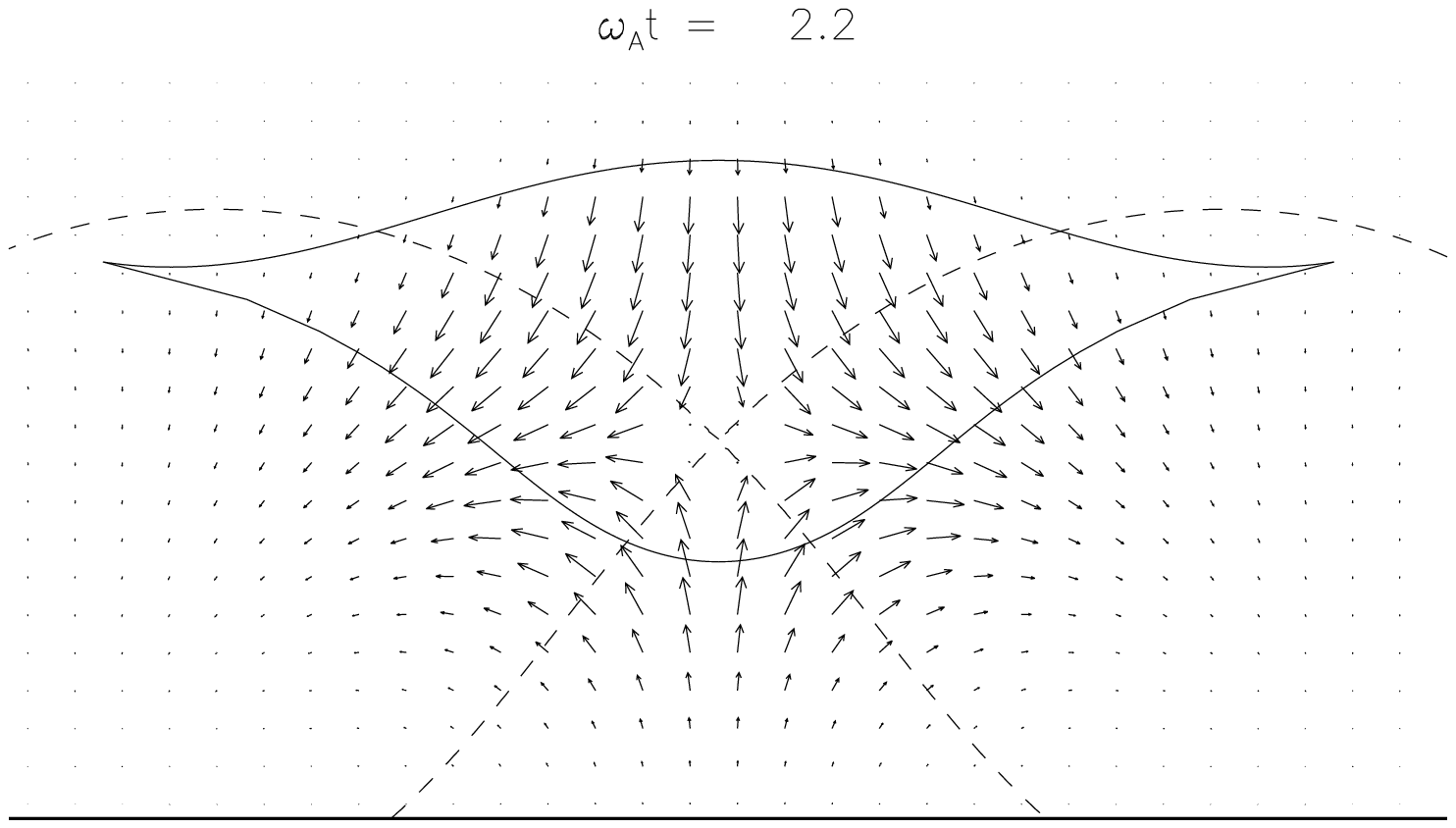}}
\caption{Velocity vectors corresponding to the electric field distribution, $U_1(y,z)$, at two early times in the solution.  Arrows show the direction and magnitude of $\vvec_1$, although their length is capped at a moderate value for clarity.  Dashed lines show the separatrices of the potential field and the solid curves are the wave fronts from the WKB approximation.}
	\label{fig:vel_plot}
\end{figure}

The inductive electric field $U_1=\vvec_1\cdot(\xhat\times\bvec_0)$ is the velocity perpendicular to qudrupolar potential field $\bvec_0$.  An axisymmetric ($m=0$) distribution of $U_1$, such as the one predicted by \citet{Longcope2007e}, corresponds to a quadruplar velocity field.  Figure \ref{fig:vel_plot} shows the velocity vectors at two times in the solution.  The plot assumes $I_{\rm cs}>0$, so the initial current sheet was horizontal like the one shown in \fig\ \ref{fig:qpl}.
(Had $I_{\rm cs}<0$, the current sheet would be vertical, $U_1<0$, and all the arrows would be reversed.)  The quadrupolar velocity field is oriented in the same way as in steady-state reconnection: inward above and below the current sheet and outward from its ends.  This sense eliminates the initial counter-clockwise azimuthal magnetic field perturbation
$\bvec_1=\nabla A_1\times\xhat$, thereby decreasing $\bvec=\bvec_0+\bvec_1$ in the inflow regions (they are rarefactions) and increasing it in the outflow regions (they are compressions).  Unlike steady reconnection models, the inflow is not here imposed from the boundary; rather it arises from the rarefaction portion of the propagating FMW.

\subsection{WKB solution}

\begin{figure}[htb]
\epsscale{1.2}
\centerline{\plottwo{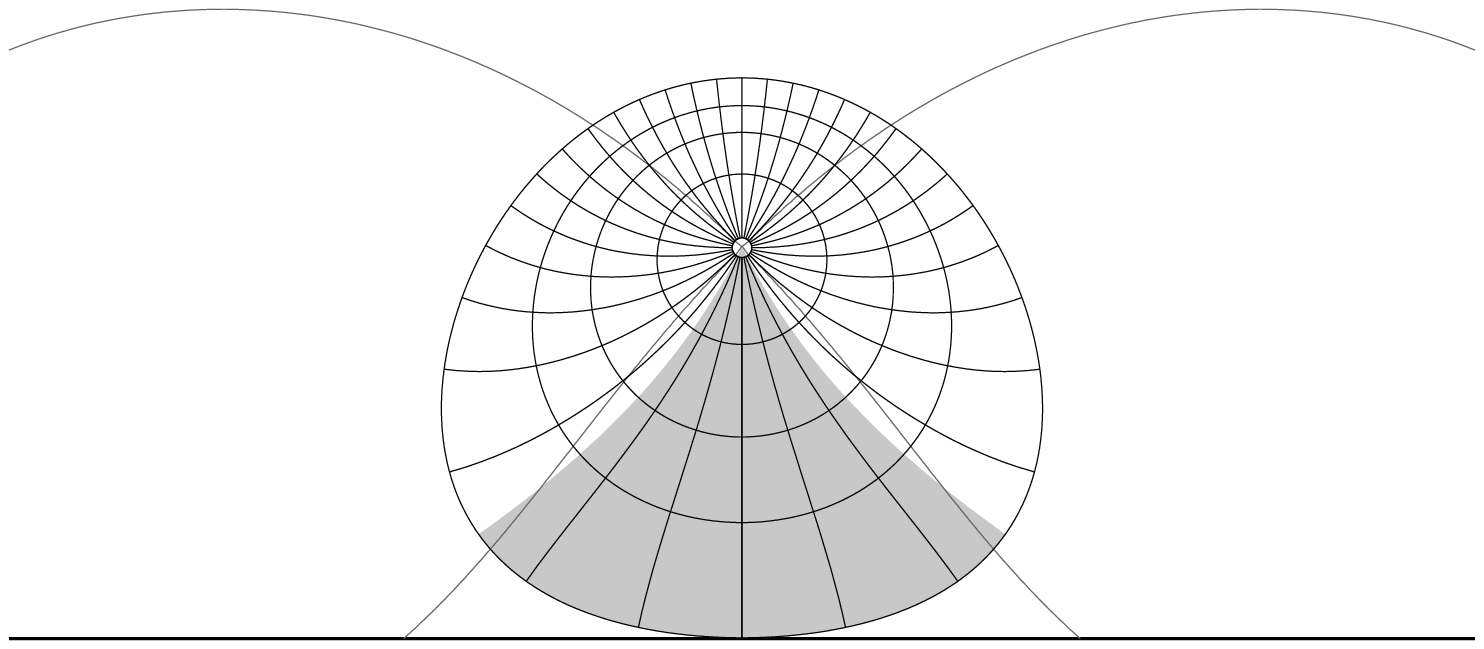}{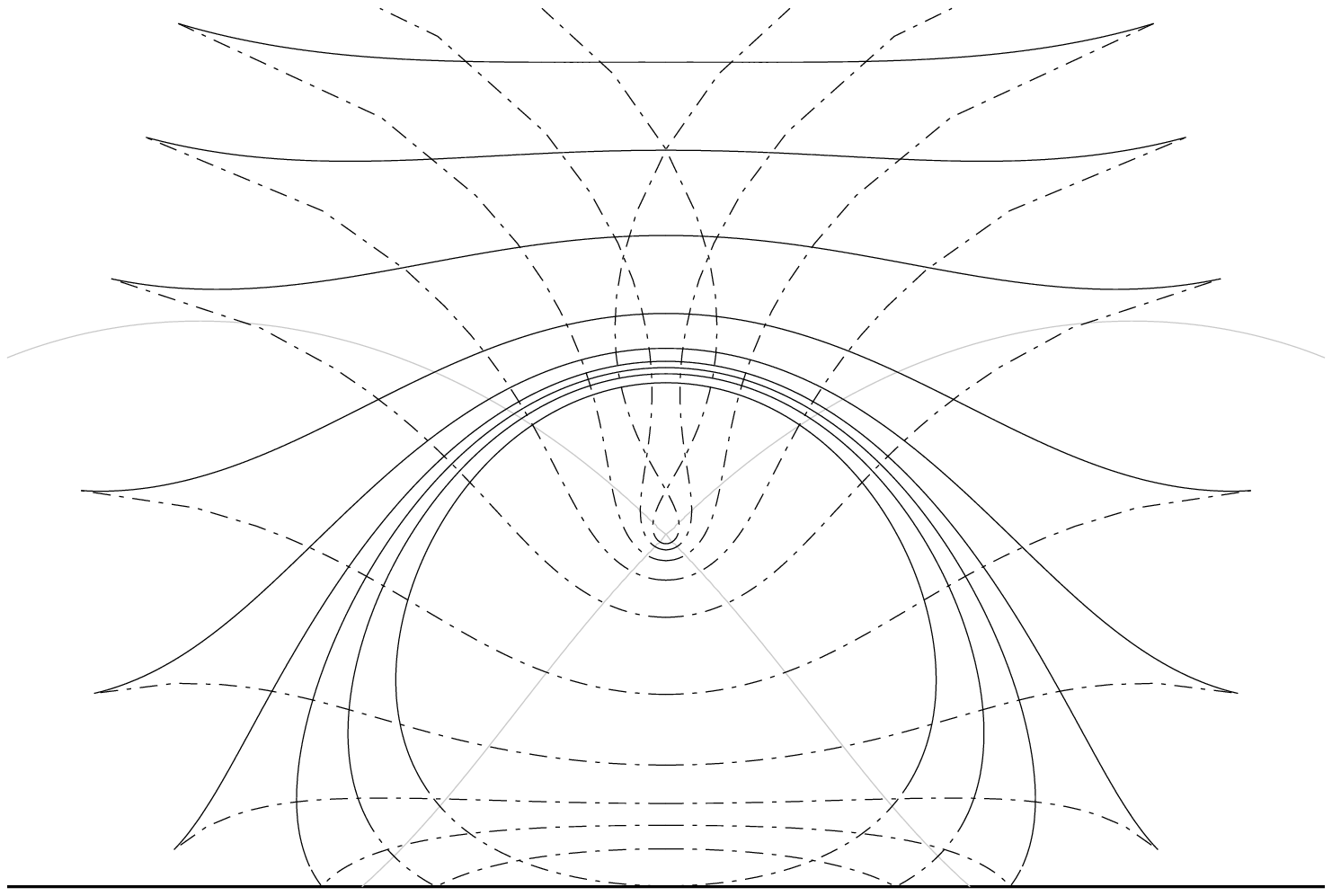}}
\centerline{\plottwo{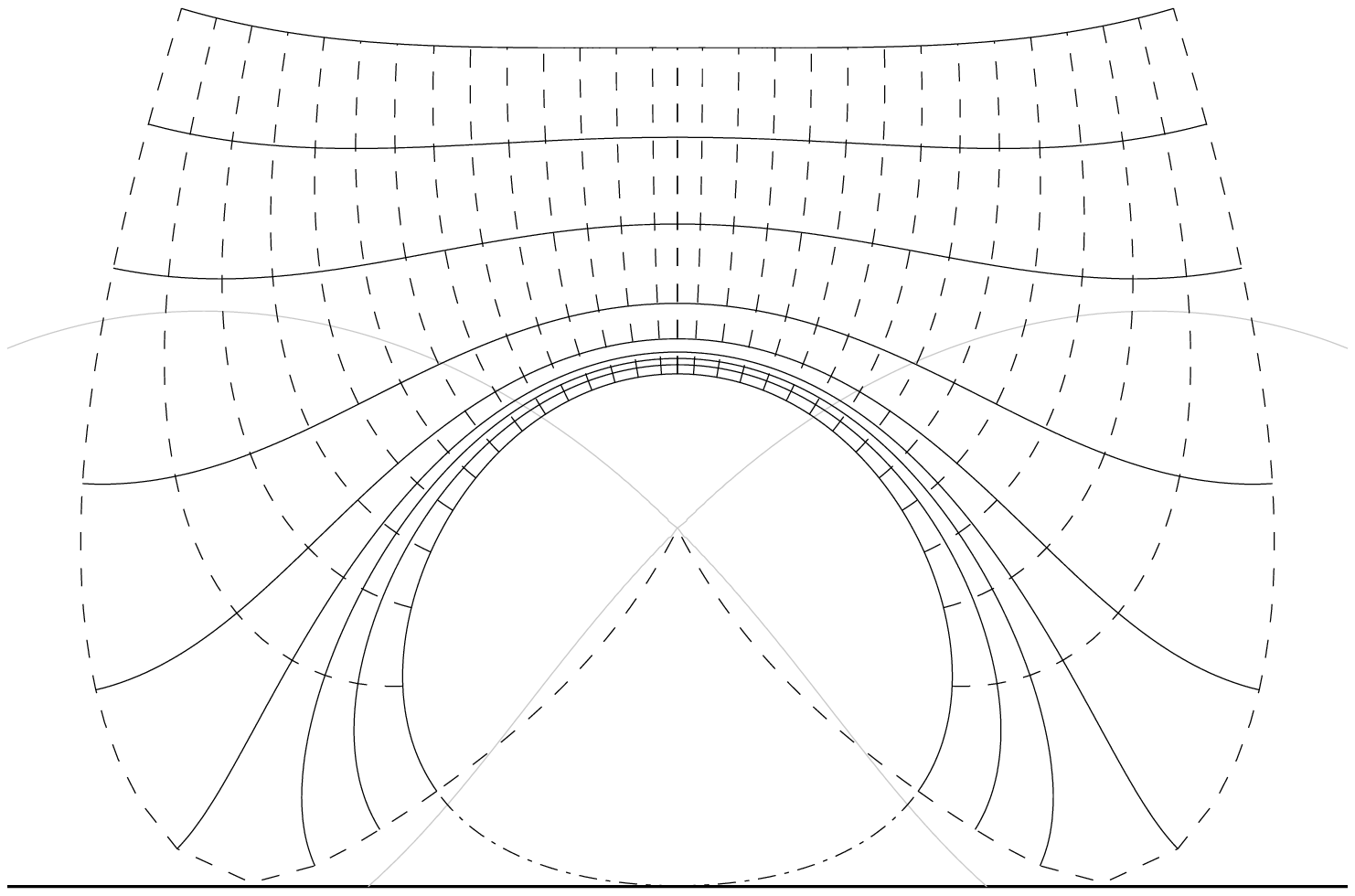}{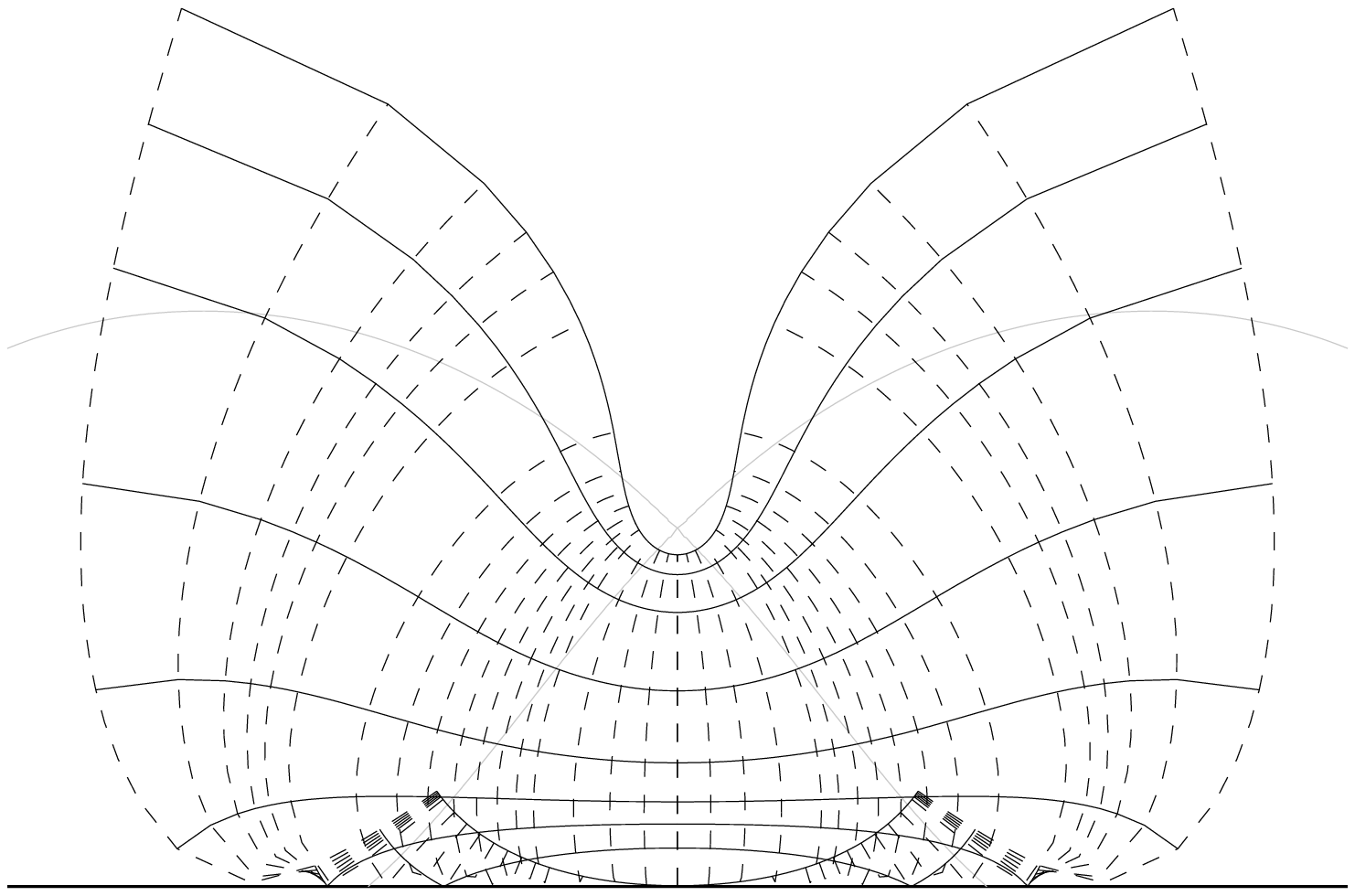}}
\caption{Solutions to the WKB equations, \eqs\ (\ref{eq:Ham1})--(\ref{eq:Ham2}).  The separatrices of the potential magnetic field are plotted in each panel with light grey.  (a) Rays and fronts for times 
$t\le t_{\rm tr}$, before any reflection.  The grey region includes the rays destined to reflect off the photosphere.  (b) Fronts for times $t\ge t_{\rm tr}$.  Broken lines show the sections reflected from the photosphere, solid shows the unreflected sections.  (c) Rays (dashed) and fronts (solid) from the rays reflecting off the photosphere.  (d)  Rays (dashed) and fronts (solid) from the rays not reflecting off the photosphere.}
	\label{fig:WKB}
\end{figure}

The structure of this typical solution can be understood in terms of an WKB approximation, valid for the components of highest frequency.   For the spatially varying dispersion relation, $\omega(\xvec,\kvec)=\va(\xvec)|\kvec|$, a ray path $\rvec(t)$ must satisfy Hamilton's 
equations \citep{Kulsrud2005,McLaughlin2006}
\begin{eqnarray}
  {d\rvec\over dt} &=&{\partial\omega\over\partial\kvec} ~=~ \va(\rvec)\,{\kvec\over|\kvec|} ~~,\label{eq:Ham1} \\
  {d\kvec\over dt} &=& -{\partial\omega\over\partial\rvec} ~=~ -|\kvec|\,\nabla\va ~~.\label{eq:Ham2}
\end{eqnarray}
A ray is initialized with $\rvec(0)$ located somewhere on the circle $r=r_i$ and $\kvec$ directed radially outward (its magnitude is irrelevant).  The full ray is then found by solving \eqs\ (\ref{eq:Ham1}) and (\ref{eq:Ham2}).  If $\rvec$ encounters the $z=0$ surface the ray is specularly reflected by changing $(k_y,k_z)\to(k_y,-k_z)$ at that instant and the continuing the solution.  The upper left panel of \fig\ \ref{fig:WKB} show rays initiated at different positions along the interface circle.  It also shows nested, closed wave fronts formed by all rays at the same time $t$.  These approximate the leading edge of the $U_1(y,z)$ front, as for example in \fig\ \ref{fig:Usol}a.

It is to be expected that the rays initially directed upward will not reflect from the photosphere, and this expectation is borne out.  In fact only a fraction of those rays initially directed downward end up reflecting; these are within the shaded region of
\fig\ \ref{fig:WKB}a.  The remainder are refracted away from the large Alfv\'en speed above the sources, and propagate upward in the end (\fig\ \ref{fig:WKB}c).

The first reflection occurs for the ray propagating straight downward, $\khat=-\zhat$, after
\be
  t_{\rm tr} ~=~ \int\limits_0^{\zxpt-r_i}{dz\over \va(0,z)} ~\simeq~ {1.42\over\oma} ~~.
\ee
(This is slightly less than the time, $\ln(\zxpt/r_i)/\oma\simeq1.74/\oma$, which would be taken in a pure X-point field due the increased field strength between the inner two sources.)  The rays beginning with angles further from vertical reflect off the photosphere successively later (see \fig\ \ref{fig:WKB}d), but only out to $\sim28^{\circ}$ of downward ($\sim 15.5\%$ of all rays).  The ray bounding the reflected set is one that appears to graze the $z=0$ surface tangentially, and is therefore marginally reflected ($k_z=0$ at the point of grazing, so its unimpeded motion is equivalent to specular reflection.)  In this case the ray grazes a point between the photospheric sources (compare its location to the separatrices shown in lighter lines).  The ray is then refracted upward by the large 
Alfv\'en speed of the outer source.

The reflected rays are sent upward from $z=0$, but only a fraction of these appear to focus back on the X-point.  In the vicinity of the X-point, where $\va(\rvec)\simeq\oma r$, Hamilton's equations are solved by rays forming logarithmic spirals
\citep{McLaughlin2006},
\be
  r(t) ~\simeq~ r_i\,e^{\oma(t-t_i)\cos\theta_0} ~~~~,~~~~\phi(t) ~=~\phi_i + \oma(t-t_i)\sin\theta_0 ~~,
\ee
where $\theta_0=\tan^{-1}(k_{\phi}/k_r)$ is the angle of the ray relative to radial, and $t_i$ is the time it crosses $r=r_i$ at angle $\phi_i$.    Of the reflected rays, only those rays for which $k_r<0$, and thus $\cos\theta_0<0$, will approach the X-point.  The wave-front from these rays drapes around the X-point, later wrapping about it as the rays complete spiral circuits.  Initial draping is evident in \fig\ \ref{fig:WKB}b.

\begin{figure}[htb]
\epsscale{1.2}
\centerline{\plottwo{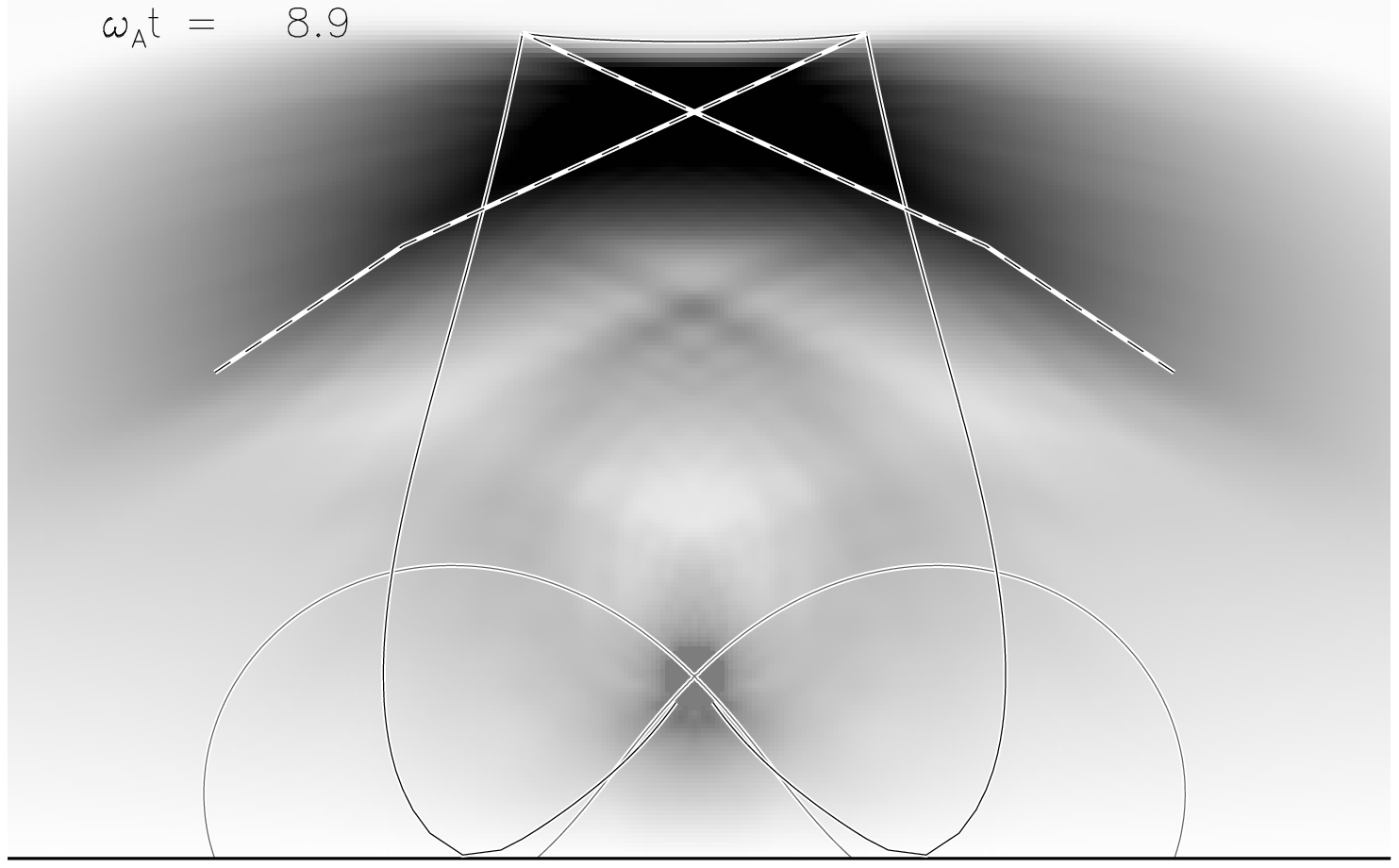}{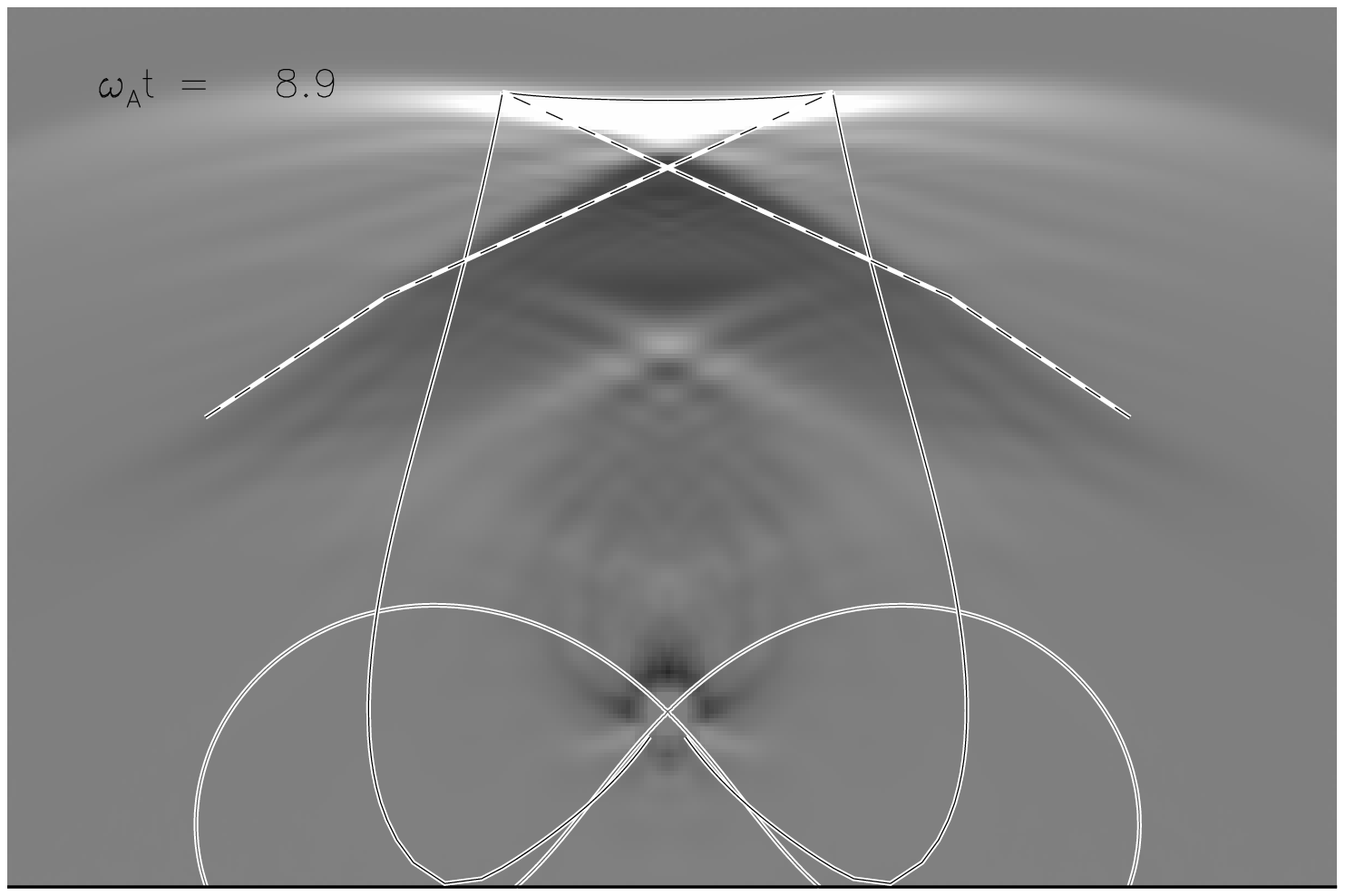}}
\caption{The solution at a time later than those shown in \fig\ \ref{fig:Usol}.  The left panel shows $U_1(y,z)$ in an inverse grey scale, but expanded by a factor of 3.  Superimposed on this are the wavefront, from the same time,of all non-reflected rays (horizontal solid segment) as well as the grazing rays (vertical solid segments).  Dashed segments are a small section of the wave  front from reflected waves.  The right panel shows the current density $J_z$ from the same time, with the same superimposed WKB solution.}
	\label{fig:late}
\end{figure}

The set of non-reflected rays, bounded by the grazing rays, forms a locus of rays bound upward.  This WKB construction is reflected in the structure of the full solution at late times, as shown in \fig\ \ref{fig:late}.  A concentrated pulse of inductive electric field, $U_1$, approximately coincides with the non-reflected rays.  The wave-front of the reflected rays adjacent to the grazing ray (dashed curve), forms an approximate lower bound to this region.  At this stage the inward flow has become decoupled from the X-point, so it is no less clearly related to the reconnection inflow.  The current density, $J_x=-\nabla^2A_1/4\pi$, shown in \fig\ \ref{fig:late}b, is also concentrated within this region.  Below the reflected front the current density is negative, while above it the current has the same sign as the current carried by the sheet: $I_{\rm cs}>0$.  The total current within the positive concentration slightly exceeds, that from the initial sheet.  

\subsection{Reflected waves}

The WKB solution shows that only a small fraction of ray paths ($15.5\%$ in this case) leaving the X-point vicinity reflect from the photosphere.  Of these an even smaller fraction converge back on the X-point.  All but one of these rays approach the X-point in a spiral, and therefore approach more slowly than they left.  We therefore expect the reflected wave incident on the X-point, $W_-$, to be of lower frequency than that emitted, $W_+$.  The WKB approximation is, however, applicable only to very high frequencies.  We therefore look to the full numerical solution to understand the waves incident on the inner radius $r_i$.

\begin{figure}[htbp]
\epsscale{1.2}
\centerline{\plotone{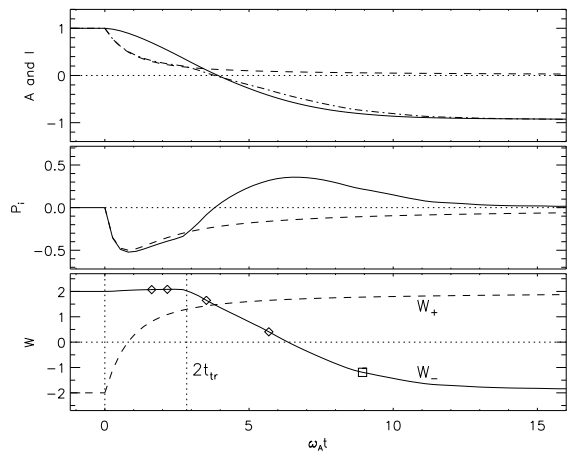}}
\caption{The waves at the interface from the solution shown in \figs\ \ref{fig:Usol} and \ref{fig:late}.  The lowest panel shows $W_-$ (solid) and $W_+$ (dashed), normalized to $2\oma I_{\rm cs}$.  Diamonds show the times of the the four panels in \fig\ \ref{fig:Usol} and a square shows the time of \fig\ \ref{fig:late}.  Vertical dotted lines show the time of initiation, $t=0$, and first reflection, $t=2t_{\rm tr}$.  The middle panel shows the outward energy flux crossing the $r=r_i$ boundary, computed using \eq\ (\ref{eq:P0}) and scaled to $\oma I_{\rm cs}^2/2$. The dashed curve shows the version for purely outging wave, plotted in \fig\ \ref{fig:LP07}.  The top panel shows the inner current, $I_i$ (broken), and flux function, $A_1$ (solid), both scaled to their values at $r=r_i$.  A dashed curve shows the current for the purely-outgoing case.}
	\label{fig:waves}
\end{figure}

Figure \ref{fig:waves} shows the incident wave, computed from \eq\ (\ref{eq:W_inc}), in the same solution from \figs\ \ref{fig:Usol} and \ref{fig:late}.  It begins to depart from the initial value, $W_-=2\oma I_{\rm cs}$, at the time $2t_{\rm tr}$, taken for a vertical ray to make a complete round-trip.  Its evolution following that is on time scales longer than the initiating wave ($W_+$ shown dashed), as predicted by the WKB solution.  

While the high-frequency components are only partially reflected, it seems the lowest frequency components are more completely reflected: $W_-\to -W_+$ by the end of the solution.  This was predictable from the outset.  In order to sustain the initial equilibrium, where $W_-=-W_+$, it is necessary for the photosphere to be a perfect inverting 
reflector at $\omega=0$, so $\hat{R}_o(0)=-1$.  Nor can this result from the outer computational boundaries, since a current sheet equilibrium is achievable, in principle, in an unbounded half-space.  

As a consequence of the high reflectivity at low frequencies, the Poynting flux through the interface (middle panel) approaches zero.  The net current inside the inner region,
\be
  I_i ~=~ -{1\over 2}r_i\left.{\partial A_1\over\partial r}\right\vert_{r_i} ~=~ {1\over 4\oma}(W_--W_+) ~~,
\ee
reverse sign and achieves a magnitude close to its initial: $I_i\to -I_{\rm cs}$.  This same behavior was observed in the case of a perfect cylindrical reflector, where it was found to eliminate and then reverse the residual X-point current \citep{Longcope2007e}.  It would establish a new vertical current sheet in place of the old horizontal one, if there were not resistivity in the X-point vicinity.  The reversed current also leads to a reversal of the flux function, $A_1(r_i)$.  This tracks the current $I_i$, as predicted by equilibrium \eq\ (\ref{eq:A_out}), except for the dynamical effects of waves crossing the interface.

\begin{figure}[htb]
\epsscale{1.2}
\centerline{\plotone{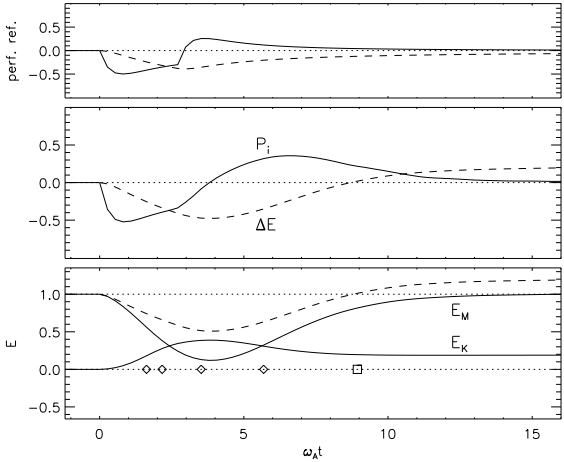}}
\caption{The energetics of the external dynamics.  Solid curves in the bottom panel shows the kinetic (lower) and magnetic (top) energies, scaled to $\Delta E_M^{\rm (out)}$.  Diamonds show the times of the the four panels in \fig\ \ref{fig:Usol} and a square shows the time of \fig\ \ref{fig:late}.  The dashed curve is the total energy.  The middle panel shows the Poynting flux (solid, scaled to $2\oma I_{\rm cs}^2$) and its integral (dashed, scaled to $\Delta E_M^{\rm (out)}$).  The top panel shows the same quantities computed for the case of perfect reflection from a concentric cylindrical conductor.}
	\label{fig:ergs}
\end{figure}

Based on the current $I_i$ it would seem that the external system has been returned to its initial state; this is the case for the perfect reflection from a conducting cylinder.   The free magnetic energy and kinetic energy of the external solution are
\begin{eqnarray}
  E_M &=& {1\over8\pi}\int\limits_{r>r_i} |\nabla A_1|^2\, dy\, dz ~~,\\
  E_K &=& {1\over 8\pi}\int\limits_{r>r_i} {U_1^2\over\va^2}\, dy\, dz ~~.
\end{eqnarray}
The energies from the foregoing solution are plotted in \fig\ \ref{fig:ergs}, along with the total energy, $E_M+E_K$ (dashed).  The initial magnetic energy equals the value from \eq\ (\ref{eq:DW_out}), and all energies are scaled to this.    Since resistivity is dropped from the external dynamics, the total energy changes only by energy flux, $P_i$, across the inner boundary.  (Since $U_1=0$ on all outer boundaries there is no Poynting flux across them.)  

As the FMW passes into the exterior region, it draws energy inward, driving the total external energy down.  The kinetic energy increases, but the magnetic energy decreases more rapidly.  The incidence of the reflected wave, $W_-$, on the interface reverses the sense of energy flux, since it causes $|W_-|$ to decrease.  At the instant $|W_-|=|W_+|$, the sense of energy flow across the interface changes from inward to outward.  Owing to the different time scales of the outward wave and reflected wave, the positive energy flux lasts longer than the negative.  The result is that the total energy of the external solution {\em increases} by $\sim20\%$ of its initial value, by the time the reflected wave is fully absorbed.  

The energy increase contrasts with the cases of perfect reflections, previously studied (shown in the top panel), where the outgoing and reflected time-scales match and the net energy flux approaches zero.  In the more realistic case considered here the state following absorption of the reflected wave includes an upward-propagating FMW  (see \fig\ \ref{fig:late}) whose kinetic energy remains in the solution.  At the time of \fig\ \ref{fig:late}, the kinetic energy density, $U_1^2/\va^2$, is evidently concentrated within the upward pulse confined to the WKB envelope.  It seems that the slight reduction in equilibrium magnetic energy accompanying the slight reduction in current ($|I_i|<I_{\rm cs}$), is roughly compensated by the magnetic perturbation of the FMW.

This corroborates our initial expectation that imperfect reflection from the photosphere would reduce the ability of the X-point to directly dissipate energy.  After a single reflection, the FMW has extracted $\sim 0.2\,\Delta E_M^{\rm (out)}$ 
from the interior current sheet.  This is energy not available for direct dissipation, later, at the current sheet.  It is the first part of the total energy which will be radiated by fast magnetic reconnection.  To estimate the effect of repeated reflections, and thereby estimate the total energy radiated in FMWs, we turn to the reflection coefficient 
$\hat{R}_o$, defined in \eq\ (\ref{eq:Ro}).

\section{Long-time solution: repeated reflections}
\label{sec:long_time}

\subsection{Repeated reflections}

The analysis above showed how the energetics of reconnection could be computed using only the amplitudes of incoming and outgoing waves at the interface $r=r_i$.  These waves satisfy linear equations so they can be computed from a superposition of successive reflections
\be
  W_{\pm} ~=~ W^{(0)}_{\pm} ~+~  W^{(1)}_{\pm} ~+~ W^{(2)}_{\pm} ~+~\cdots ~~.
  	\label{eq:Wsum}
\ee
The leading terms, $W_{\pm}^{(0)}$, are the solutions computed in the previous section: $W_+^{(0)}$ is the wave from the current disruption given by (\ref{eq:W_in}), and and $W_-^{(0)}$ is its reflection from the external solution shown in \fig\ \ref{fig:waves}.   The initial wave form, $W_+^{(0)}$, is valid only when the incident wave is steady: $W_-=2I_{\rm cs}$.  We thus define the next correction, $W_+^{(1)}$, to be the response to the departure of the incident wave from this, $W_-^{(0)}-2I_{\rm cs}$.  The Fourier transform of this departure is $\hat{W}_-^{(0)}-4\pi I_{\rm cs}\delta(\omega)$, and its reflection from the X-point is
\begin{eqnarray}
  \hat{W}_+^{(1)}(\omega) &=& \hat{R}_i(\omega)\,\Bigl[ \,\hat{W}_-^{(0)}(\omega) \,-\, 4\pi I_{\rm cs}\,\delta(\omega)\,
  \Bigr] \nonumber \\
  &=& \hat{R}_i(\omega)\,\hat{W}_-^{(0)}(\omega) \,-\, 4\pi I_{\rm cs}\,\delta(\omega) ~~,
\end{eqnarray}
after using the fact that $\hat{R}_i(0)=1$.  This correction wave then reflects from the exterior to provide the correction to the incident wave $\hat{W}_-^{(1)}=\hat{R}_o\hat{W}_+^{(1)}$.  The remainder of the terms are subsequent reflections of the correction
\be
  \hat{W}_+^{(n)} ~=~ \hat{R}_i\, \hat{W}_-^{(n-1)} ~,~ \hat{W}_-^{(n)} ~=~ \hat{R}_o\, \hat{W}_+^{(n)} 
  ~,~n\ge2 ~~.
\ee

The X-point reflection coefficient is given in analytic form by \eqs\ (\ref{eq:Rm}) and (\ref{eq:Ri}).  Its dependance on 
$\eta$, through the phase-delay factor in \eq\  (\ref{eq:Ri}), is the only place where resistivity enters the computation.  The coefficient of outer reflection, $\hat{R}_o(\omega)$, can be found from the numerical solution of a single reflection, such as the one from the previous section, by taking the ratio of Fourier transforms $\hat{W}_+$ and $\hat{W}_-$.  It proves to be more efficient to perform the computation with a different input wave form, $W_+$, whose power is more evenly distributed over frequencies.  For this we use a Gaussian $W_+(\oma t-\ln r_i) = \exp[-t^2/\tau^2]$.  Taking $\tau$ somewhat larger than $1/\oma$ samples a broader range of frequencies, and more equitably, than the smoothed-out step of \eq\  (\ref{eq:W_in}).  We solve for the linear response as before, except the initial condition is $A_1=0$.

\begin{figure}[htb]
\epsscale{1.2}
\centerline{\plotone{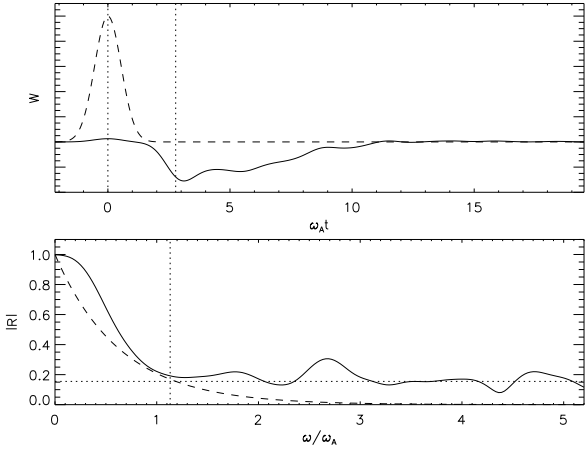}}
\caption{The external response to a Gaussian wave.  The top panel shows the input, $W_+$, a Gaussian (dashed) and the response $W_-$ (solid).  The vertical dotted lines show $t=0$ and $t=2 t_{\rm tr}$.  The bottom panel shows the magnitude of the reflection coefficient $|\hat{R}_o|$ (solid), and the analytic form of the X-point reflection, $|\hat{R}_i|$ (dashed).  The vertical dotted line is $\omega=\pi/2t_{\rm tr}$, and the horizontal dotted line is at 15.5\%, the fraction of rays reflected in the WKB solution.}
	\label{fig:lin_resp}
\end{figure}

Figure \ref{fig:lin_resp} shows the magnitude of the reflection coefficient, $|\hat{R}_o|$, computed this way.  The reflected wave had decayed to nearly zero, and the run was stopped, before any effect had reflected from the upper ($z=L_z$) or lateral boundaries ($y=\pm L_y/2$), so the response is applicable to arbitrarily large domains.  $|\hat{R}_o|$ is slightly larger than the coefficient of the X-point, $|\hat{R}_i|$ (dashed), given in \eq\ (\ref{eq:R_mag}).  It is largest at frequencies below that from the transit time, $\omega=\pi/2t_{\rm tr}$, and approaches unity at $\omega\to0$, as it must.  At higher frequencies, $\hat{R}_o$ approaches a floor near the level of reflection predicted by the WKB calculation: 15.5\%.

The X-point reflection, $\hat{R}_i$, has a finite negative slope to the right of $\omega=0$, and by realizability conditions, a positive slope to its left.  Its first derivative is discontinuous at $\omega=0$.  Such a discontinuity is accompanied by an impulse response decaying as $t^{-1}$.  This residual response was first noted by \citet{Hassam1992}, and is related to asymptotics of \eq\ (\ref{eq:W_in}) due to the shrinking region of residual current \citep{Longcope2007e}.  The external reflection coefficient, by contrast, appears to approach $\omega=0$ with zero slope.  The assymptotics of the external region will be governed by reflection, with losses, from the photosphere.  

\begin{figure}[htb]
\epsscale{1.2}
\centerline{\plotone{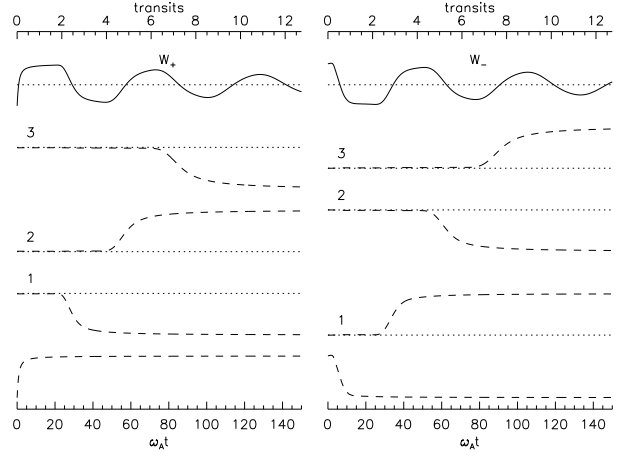}}
\caption{The different components of the reflecting waves, for Lundquist number $S=3\times10^{10}$.  Left and right panels show $W_+$ and $W_-$ respectively.  Dashed curves are individual components, $W^{(n)}_{\pm}$ for $n=0$, 1, 2, and 3 (from bottom to top).  The solid curve is the sum up to $n=14$.  The top axis shows time in units of the full transit, 
$t_{\rm tr}+\ln(r_i/\ell_{\eta})/\oma$.}
	\label{fig:long_waves}
\end{figure}

Figure \ref{fig:long_waves} illustrates this procedure for a Lunquist number 
$S=\zxpt^2/\ell_{\eta}^2=3\times10^{10}$.  This value enters only into the phase factor in \eq\ (\ref{eq:Ri}), representing the transit delay, $\ln(r_i/\ell_{\eta})/\oma=10/\oma$, between interface and resistive scale.  This plus $t_{\rm tr}$ is half the round-trip delay for reflections off both the photosphere and the X-point. It is therefore approximately half the delay between the rising edges of $W_{\pm}^{(n-1)}$ and $W_{\pm}^{(n)}$.  Each reflection is a step of opposing sign, delayed by this amount and broadened by the low-pass structure of $|\hat{R}_i|$ and $|\hat{R}_o|$.  Were it not for the broadening the composite wave (i.e.\ \eq\ [\ref{eq:Wsum}]) would be a series of steps.  Instead it becomes increasingly sinusoidal as higher frequency components are damped away at each reflection.  (The same tendency was found by \citet{Longcope2007e}, and is evident in their \fig\ 11.)

\begin{figure}[htb]
\epsscale{1.2}
\centerline{\plotone{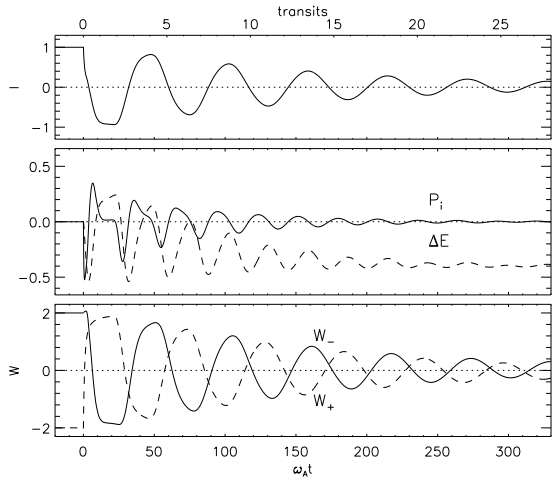}}
\caption{The energy flux across the interface from the waves in \fig\ \ref{fig:long_waves}.  The bottom panel shows $W_+$ (dashed) and $W_-$ (solid); this extends, through independent computation, the bottom panel of \fig\ \ref{fig:waves}.  The middle panel shows the Poynting flux, $P_i$  and its integral, $\Delta E$ in the same format as in \fig\ \ref{fig:ergs}.  The top plot is current $I_i$.}
	\label{fig:long_erg}
\end{figure}

The energy flux due to this wave, plotted in \fig\ \ref{fig:long_erg}, shows the initial behavior found in \fig\ \ref{fig:ergs}: a brief positive flux, followed by a longer negative (inward) flux.  This repeats once more when the wave reflects from the X-point, this time with a longer negative portion and shorter positive portion, leading to less net outward flux.  Thereafter the waves become increasingly sinusoidal assuming a phase relation where the upward zero crossing of  
$W_+$ is slightly ahead of the downward crossing of $W_-$; the ratio is $\sim t_{\rm tr}\oma/\ln(r_i/\ell_{\eta})$, due to the difference in transits.  This phasing leads to a steady decrease in the net energy transport until it converges on a 
value $\sim-0.41\Delta E_M^{\rm (out)}$.  Free magnetic energy from the external field is, therefore, moved into the inner region and dissipated.  The remaining free energy, $\sim0.59\Delta E_M^{\rm (out)}$, is therefore converted into FMWs propagating away from the reconnection.  The full solution (i.e.\ \fig\ \ref{fig:Usol}), covering only the first oscillation, showed a single wave accounting for about a third of the total.  It seems that new waves are launched under repeated reflections, each with smaller amplitude.  The total energy in all waves is the 
$\sim0.59\Delta E_M^{\rm (out)}$ carried away from the exterior.

\subsection{Scaling with Lundquist number and free energy}

The external solution, leading to $\hat{R}_o(\omega)$, does not depend on the resistivity.  The inner reflection, $\hat{R}_i$, involves resistivity, through $\ell_{\eta}=\sqrt{\eta/\oma}$, in the phase-factor 
$e^{2i(\omega/\oma)\ln(r_i/\ell_{\eta})}$ in \eq\ (\ref{eq:Ri}).  It is therefore easy to perform the same calculation for any value of $\eta$.  The results all resemble
 \fig\ \ref{fig:long_erg} with predicable differences.  As $\eta$ decreases the transit time, $\ln(r_i/\ell_{\eta})$, increases and the oscillation period increases.  Successive reflections, $W_{\pm}^{(n-1)}$ and $W_{\pm}^{(n)}$, are spaced farther apart therefore cancel each other less resulting in more gradually decreasing total waves, $W_{\pm}$.  The phase difference between these approaches $180^{\circ}$ since $t_{\rm tr}$ becomes a diminishing fraction of the whole period with smaller $\eta$.  This results in a diminishing net transport inward and thus a smaller peak values of $P_i$.  This lower mean power persists for a longer time before the component waves decay. The result is a net energy transfer which depends only very weakly on the logarithm of $S$: from $-0.50\Delta E_M^{\rm (out)}$ at $S=3\times10^5$ to 
 $-0.35\Delta E_M^{\rm (out)}$ at $S=3\times10^{12}$.  A sweep of this range is well fit by a fraction
 $-0.6+0.0076\ln S$.
 
\begin{figure}[htb]
\epsscale{1.2}
\centerline{\plottwo{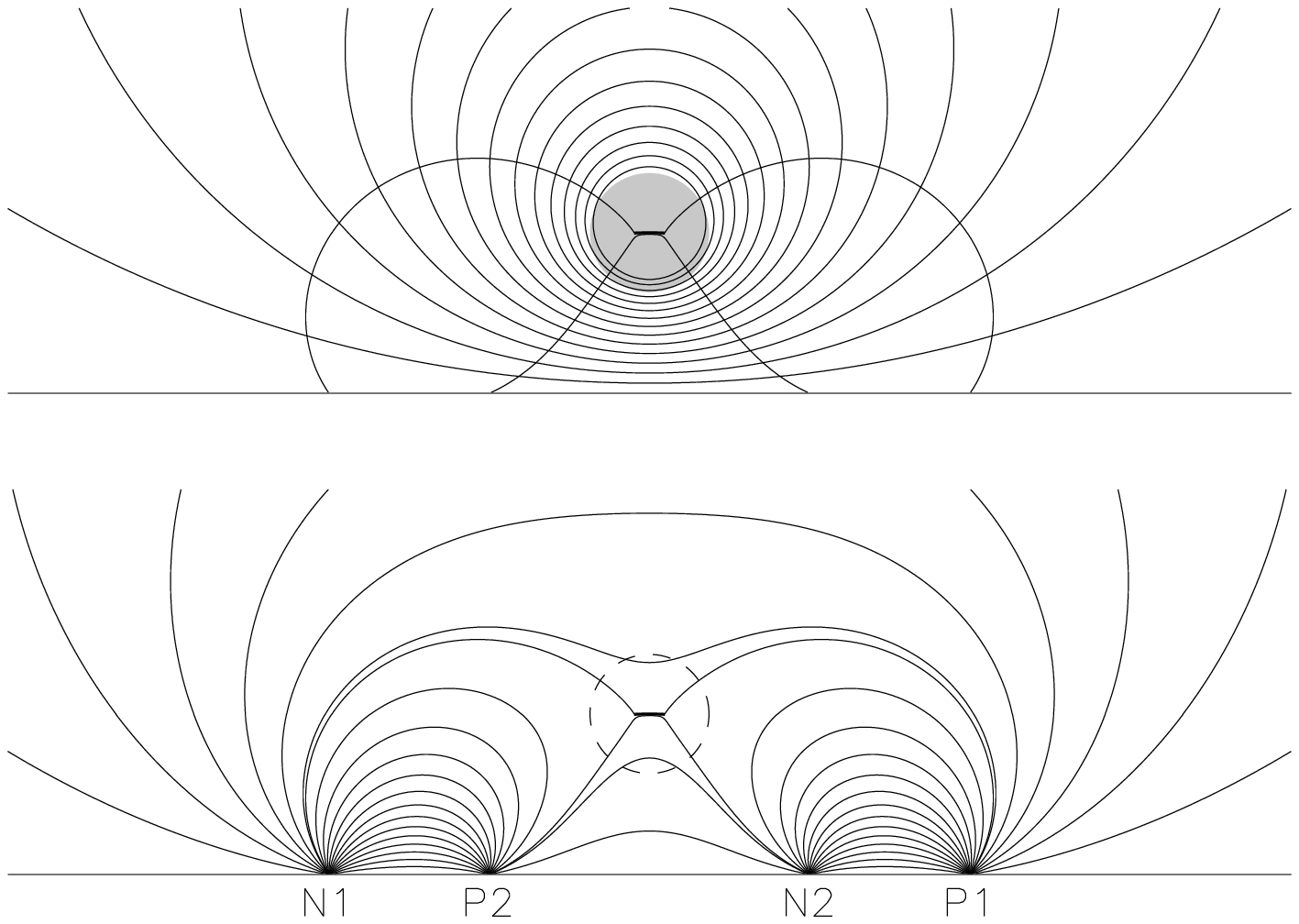}{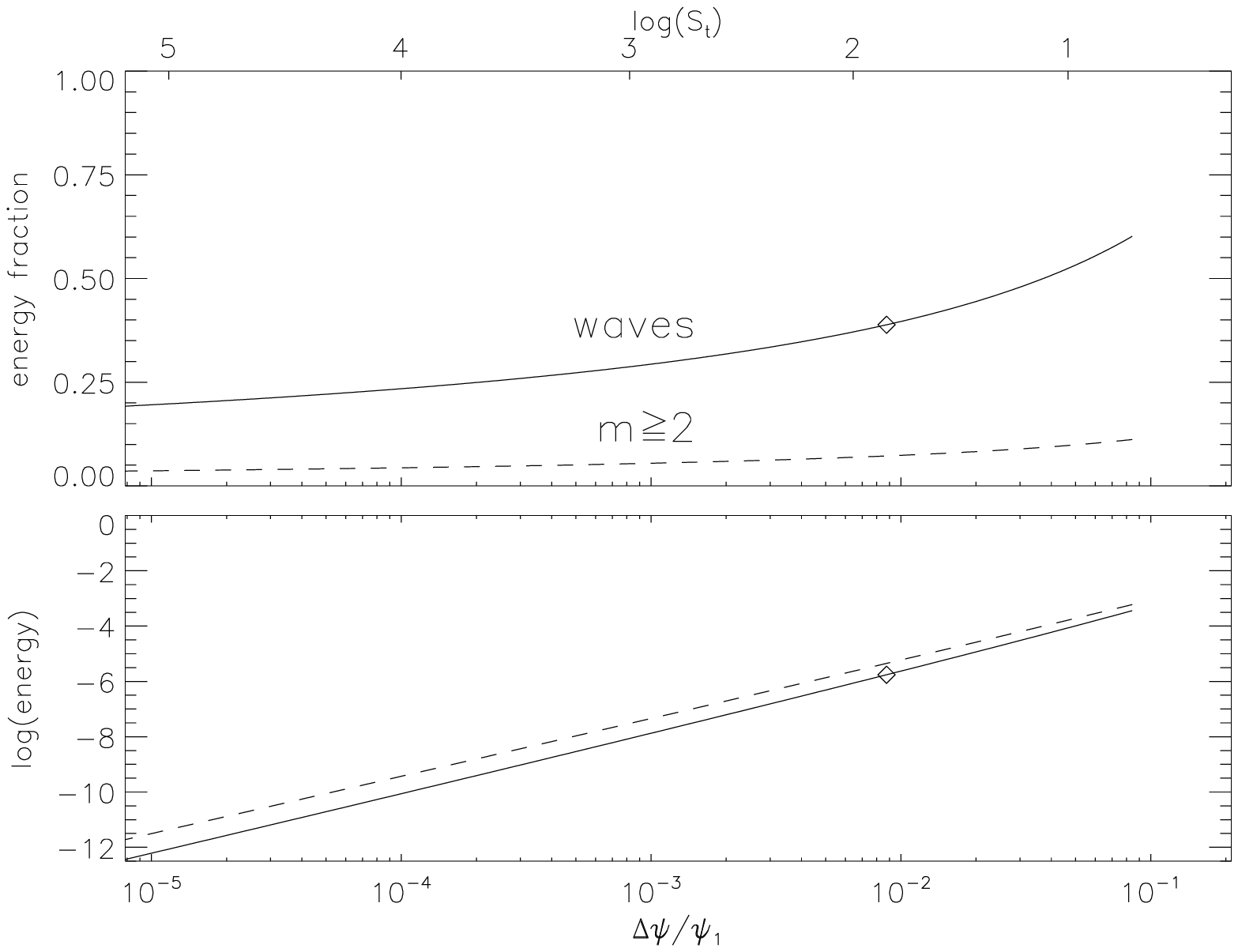}}
\caption{Left: The quadrupolar field with $y_2/y_1=-1/2$ and $\psi_2/\psi_1=4/5$, used in the forgoing calculations.  The dashed circle in the bottom panel, and the grey area in the top, define the interior region whose energy is entirely dissipated.  The energy initially outside is converted to FMWs.  The bottom panel shows the field lines, contours of $A(y,z)$, and the top shows lines of the non-potential field, contours of $A_1(y,z)$.  Right: the free magnetic energy (dashed) and energy carried by waves (solid) versus the change in flux $\Delta\psi/\psi_1$.  Plotted is the logarithm (base 10) of the energy in units of $\psi_1^2$.  The top panel shows, as a fraction of the total free energy, the energy ultimately converted to waves (solid) and the energy initially present in modes with $m\ge2$ (dashed) and then directly dissipated.  The axis along the top is the logarithm of the maximum permissible Lunduist number, from minimum permissible resistivity,
$\eta_t=\oma\Delta^2$.  Diamonds correspond to the current sheet shown in the right panel.}
	\label{fig:Xpoint_circ}
\end{figure}

We therefore find, neglecting the extremely weak dependence on $S$, that a fixed amount of magnetic energy is converted to FMWs by the reconnection.  That energy is less than all the free magnetic energy outside the radius $r_i=0.07y_1$ we used for our computation.  Using \eq\ (\ref{eq:DW_out}), we find the wave energy to be approximately equal to all of the free magnetic energy outside the larger radius $r_i\simeq0.19y_1$.  This radius is outside the validity of our calculation since $A_0(y,z)$ inside such a large $r_i$ cannot be approximated by an X-point, as in eq\ (\ref{eq:A0_X}).  We extrapolate to these values in order to illustrate the expected behavior of the system with significant free energy.  In such a case, illustrated in \fig\ \ref{fig:Xpoint_circ}, all of the free magnetic energy inside the region would be dissipated by the resistivity, all the energy outside would bet carried away by FMWs.  

This foregoing scenario appears to be virtually the same regardless of the value of resistivity responsible for the dissipation, provided it is sufficient to cause fast disruption of the current sheet.  The total free energy in the initial field depends on the length of the current sheet, $2\Delta$ (i.e.\ through \eq\ [\ref{eq:DW_tot}]), which itself depends on the current $I_{\rm cs}$.    The net current in the sheet is in turn set by the flux discrepancy, $\Delta\psi$, imposed by ideal evolution driven from the boundaries (for instance through flux emergence).  The right side of \fig\ \ref{fig:Xpoint_circ} shows the total free energy (solid line) and the amount converted to FMW; the upper panel gives the fraction of total energy converted to waves.  That fraction becomes smaller as the initial energy becomes small.  Those cases represent sheets with very small currents which are therefore themselves physically small ($\Delta \ll y_1$).  The free energy becomes increasingly concentrated in the small neighborhood of the current sheet, where it is more easily dissipated.  So dissipation eliminates a larger fraction of the energy, which is itself smaller due to the smaller current.

The foregoing analysis considered only axisymmetric waves, $m=0$, in the vicinity of the X-point.  The initial current sheet also included modes with $m\ge2$, but it is shown in the appendix that the energy in these modes is dissipated immediately by the resistive diffusion.  The energy in these modes, $\simeq0.27\,I_{\rm cs}^2$, is included in 
$\Delta E_M$, of \eq\ (\ref{eq:DW_tot}), and provides a lower limit on the energy directly dissipated by Ohmic diffusion in our model.  Plotted as a dashed curve in the upper right panel of \fig\ \ref{fig:Xpoint_circ}, it accounts for between 3\% and 11\% of the total.

Releasing significant free energy requires that the initial current sheet have non-trivial physical extent.  In this case our uniform resistivity, $\eta$, must be interpreted as the result of complex (turbulent) small-scale physics occurring rapidly within the current sheet.  In order to simplify our analysis we have assumed this turbulent diffusion is uniform in space and constant in time, following the onset of the turbulence.  It must also be large enough to disrupt the entire current sheet over a time comparable to the Alfv\'en transit across the sheet: $\eta_t \gg \oma\Delta^2$, which is equivalent to our earlier assumption that $\ell_{\eta}\gg\Delta$.  This requirement places an upper limit on the turbulent Lunquist number --- a limit which depends on $I_{\rm cs}$ through $\Delta$.  The logarithm of this limiting value is plotted across the top right of  \fig\ \ref{fig:Xpoint_circ}.  It is evidently necessary to have $S\sim100$ in order to rapidly reconnect one percent of the photospheric flux; a current sheet like that in the left panel of \fig\ \ref{fig:Xpoint_circ}.  In this case 40\% of the free energy released would be converted into FMW, and the remainder would be dissipated by the turbulent processes at the X-point.

\section{Discussion}

We have used a simplified model to study the response of the large-scale coronal magnetic field to reconnection at a current sheet.  The reconnection reduces (almost to zero) the current in the sheet by transferring magnetic flux across it.  As reported by \citet{Longcope2007e}, the current removed from the sheet is carried away at the front of a fast magnetosonic wave.  A portion of this wave reflects from the photospheric boundary, and is refracted back toward the X-point.  Subsequent reflections of the wave between the X-point and the photosphere lead eventually to the elimination of all current, and with it all free magnetic energy.  

The number of reflections required to reach the potential field is roughly the logarithm of the global Lundquist number, corroborating the findings of \citet{Craig1991} and \citet{Hassam1992}.  The total reconnection time is therefore 
$\sim\ln^2 S$ times a single Alfv\'en transit (i.e.\ $1/\oma$).  This final approach resembles the case of perfectly reflecting concentric cylindrical boundary because the photospheric boundary has a much higher reflection coefficient than the resistive X-point at very low frequencies.  During early reflections, however, the FMW contains frequencies above $\oma/\ln S$, of which a sizable fraction are directed vertically upward from the photosphere.  We thus find, at least for this simplified model, that an appreciable fraction of the initial free energy is carried from the current by fast magnetosonic waves; in the case we explored in detail 25\% -- 60\% of the energy was converted to waves, depending on the size of the initial current sheet.

We expect some of the basic elements of this result to hold in models more sophisticated that the one we used for our detailed study.  A common element in all models of fast magnetic reconnection is that the reconnection electric field, i.e.\ the flux transfer, occurs on small scales, while free magnetic energy is stored over vastly larger scales.  Information about the flux transfer, and associated current reduction, must be transmitted to the larger corona.  This includes significant volumes which  are not magnetically linked to the reconnection site, for which transmission must occur through fast magnetosonic modes.  Previous studies of unsteady fast reconnection have shown fast magnetosonic waves propagating into the unreconnected flux to be responsible for creating the reconnection inflow, the same inflow assumed as a boundary condition or driver in steady models \citep{Lin1994,Heyn1996,Nitta2001}.  Our model contains similar fast mode rarefaction waves, and reveals the fraction of energy they contain.

One of the most dubious simplifications is our use of uniform classical resistivity, $\eta$, as the means of generating a reconnection electric field.  The simple mathematical form of this effect permits a more thorough analysis than would more complex physical effects.  Since our goal was to study the response of the large-scale field, where the electric field is irrelevant, we chose the simplest possible form for the small-scale effect.  We expect that a more accurate treatment of the small scales would reveal complex flows, whose overall effect might be characterized as an effective diffusivity.  It is this kind of turbulent $\eta$ we use in applying our results to the reconnection of global currents, as in \fig\ \ref{fig:Xpoint_circ}.  Were the turbulent diffusivity computed with any degree of self-consistency, it is unlikely to remain constant through the many repeated reflections of the FMW.  Remarkably, we find that its actual value has no effect on the magnitude of the X-point reflection (see \eq\ [\ref{eq:Ri}]), critical to the ultimate energy dissipation.  Thus we expect the basic scenario revealed in our simple model would be found in more sophisticated ones.

The equations governing the large-scale response, namely \eqs\ (\ref{eq:momentum}) and (\ref{eq:induction}), are linear, purely two-dimensional (with no magnetic field in the ignorable direction; no ``guide field'') and assume zero pressure.  The assumption of linearity is the most easily justified, since the initial current sheet contributes a small correction to the field far from  itself.  \citet{Longcope2007e} analyze in detail the requirements for linearity in the internal solution, and find that it is a reasonable approximation for many times $1/\oma$, by which time the reflected wave comes into play.  

Significant pressure can alter the nature of the equilibrium from which the energy must be released.   A recent numerical solution found that only a tiny fraction ($\sim3\times10^{-9}$) of the magnetic energy in a pressure-dominated equilibrium was released by suddenly enhancing the resistivity \citep{Fuentes-Fernandez2012}.  They found that resistive diffusion does redistribute the current, as in the $\beta=0$ case, but that the new distribution of Lorentz forces are compensated by a new distribution of pressure.  This redistribution process is localized to the current sheet so that instead of an axisymmetric FMW reducing the free energy, a small wave with $m=4$ is launched, with negligible energetic consequence.    A still more recent experiment at lower values of $\beta$ continued to observe reconnection leading to pressure re-distribution rather than a decrease in the sheet's current \citep{Fuentes-Fernandez2012b}

At the other end of the process, \citet{McLaughlin2006b}, found that pressure in the vicinity of the null kept the fast magnetosonic speed above zero, thus mitigating the focusing of wave energy there.  Pressure thus reduces the amount of energy initially released and the fraction of that dissipated.  By focusing on the zero-pressure limit we are finding the maximum possible energetic effect in solar flares, which pressure would almost certainly reduce.

We believe our assumption of two-dimensionality places the most severe limitations on the applicability of our model.  In the absence of a guide field ($B_x$) all shear Alfv\'en modes are polarized in the ignorable direction ($\xhat$) and are completely uncoupled from the reconnection dynamics.  It is for this reason that the response propagates isotropically from the X-point as a FMW.  Adding any amount of an ignorable component to the system will result in a global response including both FMWs, propagating in all directions, and shear Alfv\'en waves confined to the separatrix field lines.  While the FMW will reflect from the photosphere specularly, as they do in our model, the reflected Alfv\'en waves will travel back along the same field lines to reach the X-point again.  This far more complex, and more interesting, system will need to be studied in the future.  The present study can be taken as a limiting case, and can provide an outline, for such a future investigation.  We expect the FMW component of that system to behave in a manner similar to our model, but to account for a smaller portion of the energy, since Alfv\'en waves will contain some as well.

Many past observations have provided evidence for both FMW and shear Alfv\'en modes initiated by solar flares.  Evidence of fast mode shocks are found in coronal type II radio bursts \citep{Payne-Scott1947} and chromospheric Moreton waves \citep{Moreton1960,Moreton1960b}, while shear Alfv\'en waves are observed in post-flare loop oscillations \citep{Aschwanden1999,Nakariakov1999}.  In both cases their timing strongly suggests these disturbances are initiated by flares or by CMEs, although the details of the initiation remain unclear.  The wave front in our model includes a ``skirt'' along the lower boundary, evident in \fig\ \ref{fig:WKB}b, which might manifest as a Moreton wave \citep{Uchida1968}.  The upper portion of the wave, if correctly oriented, could shock to produce metric type II radio signatures.

A common model of these observed waves has been that direct energy dissipation creates a pressure 
pulse driving a FMW outward in all directions \citep[see][and references therein]{Vrsnak2008}.  
Our model provides a detailed picture of the initiation mechanism which differs in several important respects form pressure-pulse models.  Our primary conclusion is that, contrary to the common assumption, the broad initial distribution of free magnetic energy makes it impossible to create a local pressure pulse through its rapid dissipation.  We find instead that rapid diffusion results in a current redistribution whose newly unbalanced Lorentz forces produce the FMW.  We admittedly discard the pressure which would make possible the alternative wave-generation mechanism.  Were it included, however, the thermal energy could not exceed the magnetic energy directly dissipated.  We find this to be 3\%--11\% of the total, and therefore smaller than the wave energy produced through Lorentz forces.

The plasma flow direction is one potentially observable difference between our model and the pressure-pulse waves proposed previously.  The pressure pulse or piston driver leads to outward (compressive) flows in all directions, while our wave has both outward and inward (rarefaction) flows along different portions of the cylindrical front.  The outward portion is an extension of the inner reconnection outflow jet and thus related to the fast magnetosonic termination shock predicted in various models \citep{Forbes1986c}.  The inward portion is the reconnection inflow, but its speed is a function of the external field rather that the inner reconnection rate.   The relative locations of these two components depend on the orientation of the initial current sheet, and thus on the sign of $I_{\rm cs}$.  

Moreover, since all its energy is introduced at the outset, the pressure-driven wave has an amplitude that decreases with distance from the source.  In contrast, the FMW continues to draw energy from the large scale magnetic field and thus decays much more slowly; in the vicinity of the X-point its amplitude remains constant even as its net energy increases exponentially \citep{Longcope2007e}.  It remains to compare the amplitude profiles of observed FMWs to the predictions of these different models.

Finally, our model shows how  the dissipation of magnetic energy at the smallest scales, i.e.\ the current sheet itself, requires repeated reflection from distant boundaries (i.e.\ the photosphere), and is therefore not entirely local.  The initial phase of flux transfer dissipates only a very small fraction of the energy, as found by \citet{Longcope2007e}, and expected from arguments based on the finite energy density and very small volume.  Reflected waves will focus back onto the X-point due to its vanishing Alfv\'en speed.  As first predicted by \citet{Craig1991} and \citet{Hassam1992}, this focusing leads in the end to the dissipation of significant energy within the small region after many reflections.  The energy dissipation thus persists far longer than the time taken for the initial flux transfer that we call reconnection.  

It is tempting to see in this persistent dissipation a possible explanation for flare durations longer than free cooling following a single impulsive energy release \citep{Warren2006}.  Before doing so we must confront the effect, alluded to above, of three-dimensional geometry.  With a guide field component the fast magnetosonic speed no longer vanishes at the X-point, and we would expect far less of the wave energy to focus back there.  Alfv\'en waves, on the other hand, will be confined to the field lines and will thus reflect repeatedly back to the reconnection site.  This kind of wave reflection and dissipation, sometimes invoked to explain the persistent energy release in flares, is indirectly related to the fast mode version we have studied.  Genuine fast mode focusing would, however, occur at a coronal null point in a three-dimensional magnetic field, as it does in our two-dimensional version.

\bigskip

This work was supported by a grant from the NSF/DOE Plasma Sciences partnership.  Aaron Schye aided the effort with a preliminary WKB computation.  DWL thanks M.D.\ Ding and the Nanjing University, School of Astronomy and Space Science for hosting a visit during which some of this work was done.

\appendix

\section{Non-axisymmetric wave components from current sheet disruption}

\begin{figure}[htb]
\epsscale{0.7}
\centerline{\plotone{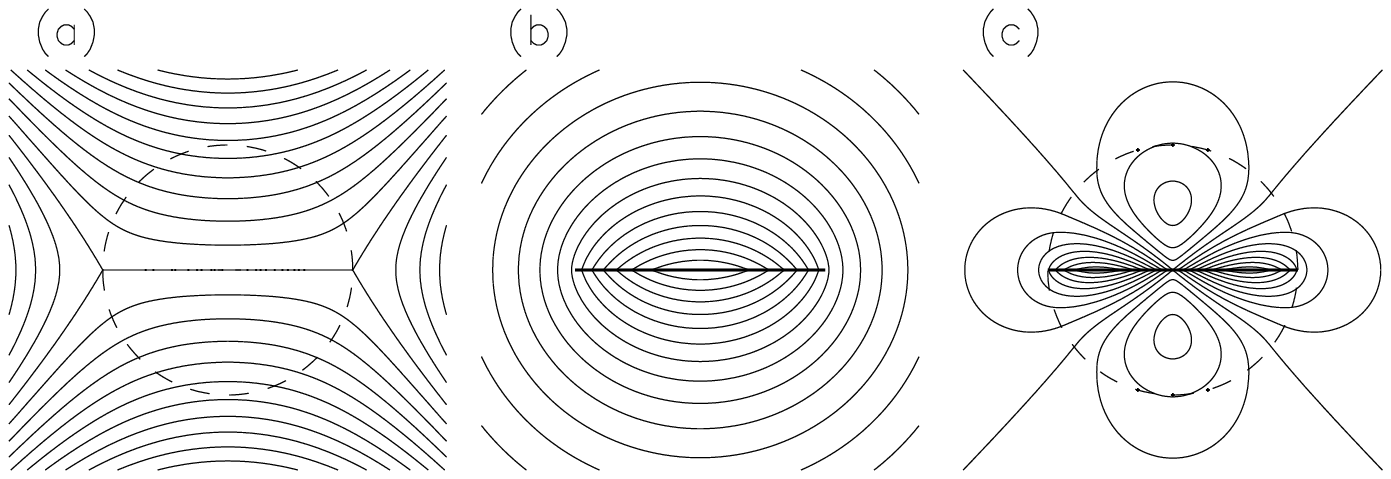}}
\caption{The field in the vicinity of the current sheet, rendered by contous of the flux function $A(y,z)$.    (a) The flux function $A(y,z)$, (b) the perturbation $A_1(y,z)$ and (c) the non-axisymmetric contribution from $m\ge2$.  The dashed circle at $r=\Delta$ divides the external expansion 
(\ref{eq:A_exp_out}) from the internal one (\ref{eq:A_exp_in})}
	\label{fig:CS}
\end{figure}

The flux function for the straight current sheet of length $2\Delta$, shown in \fig\ \ref{fig:CS}a,  can be expressed in polar coordinates as an expansion valid either for radii $r<\Delta$ or for radii $r>\Delta$.  The external expansion, for $r>\Delta$,  was given in \citep{Longcope2007e}\footnote{There was a typographical error in the $m\ge2$ terms of \eq\ (2) in \citet{Longcope2007e}, which were never used.  The expression there was missing the factor of 4 in front of the sum.  }
\be
   A(r,\phi) =\Frac{1}{2}B'_0 r^2 \cos(2\phi)-2I_{\rm cs} \ln\Bigl(2 e^{1/2}\Frac{r}{\Delta}\Bigr) 
   +4I_{\rm cs} \sum_{m=2}^{\rm evens}\Frac{(m-1)!!}{m(m+2)!!}\Bigl(\Frac{\Delta}{r}\Bigr)^m\cos(m \phi) ~~,
  	\label{eq:A_exp_out}
\ee
where $(m+2)!! =2\cdot4\cdots m\cdot(m+2)$ and $(m-1)!!=1\cdot3\cdots(m-1)$.
The first term in (\ref{eq:A_exp_out}) reproduces the simple X--point, 
while the remaining terms form the perturbation, $A_1(r,\phi)$, due to the current sheet, shown in \fig\ \ref{fig:CS}b.  
Expression (\ref{eq:A_out}) includes only the leading term in the perturbation term (axisymmetric: $m=0$), while we consider here the contributions of the remaining terms, $m\ge2$.  It is evident from \fig\ \ref{fig:CS}c that this is dominated by $m=2$, but also includes appreciable contributions from $m=4,$ 6, and so on.

The internal expansion ($r<\Delta$) of the entire flux function is  
\be
  A(r,\phi) = \Frac{1}{2}A^{(0)}(r) + \sum_{m=2}^{\rm evens}A^{(m)}(r)\cos(m\phi)
\ee
 where $m$ is once again even and
\be
  A^{(m)}(r) = \Frac{16 I_{\rm cs}}{\pi} \sum_{n=1}^{\rm odds}\Frac{n!!}{(n^2-2n)(n^2-m^2)(n-1)!!}\Bigl(\Frac{r}{\Delta}\Bigr)^n~~,
  	\label{eq:A_exp_in}
\ee
(by convention we take $0!!=1$).
While the external series involves only a single inverse even power of $r$ for each mode, the internal expansion involves all odd powers for each mode number $m$.  These decrease asymptotically as $n^{-4}$ and the series therefore converges rapidly.  
From these expressions we find the total free energy contributed by non-axisymmetric modes ($m\ge2$), the field in \fig\ \ref{fig:CS}c to be, $0.27I_{\rm cs}^2$.  This contribution is included in the total energy, \eq\ (\ref{eq:DW_tot}), but is a very small fraction provided $\Delta\ll\zxpt$.  


Starting with the linearized resistive MHD equations (\ref{eq:momentum}) and (\ref{eq:induction}), we again take a polar expansion of our perturbed flux function and electric field variable
to obtain the scalar \eqs\ (\ref{eq:U1}) and (\ref{eq:A1}), only now keeping the resistive term:
\begin{eqnarray}
  \label{eq:u1}\Frac{\partial U_1}{\partial t} &=& -\oma^2r^2\nabla^2 A_1\\
  \label{eq:a}\Frac{\partial A_1}{\partial t} &=& -U_1 + \eta \nabla^2 A_1.
\end{eqnarray}

\begin{figure}[htb]
\epsscale{0.4}
\centerline{\plotone{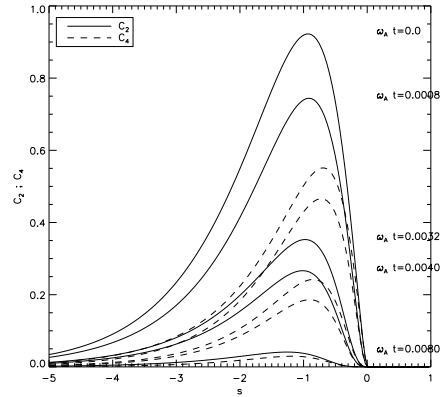}}
\caption{Evolution of $C^{(2)}$ (solid) and $C^{(4)}$. (dashed) for the case where $\eta=\oma\Delta^2$ 
(i.e.\ $\Delta=\ell_{\eta}$).  Each profile is plotted {\em vs.} $s=\ln(r/\Delta)$ for five different times, 
$\oma t=0,\,8\times10^{-4},\, 3.2\times10^{-3},\, 4\times10^{-3}$ and $8\times10^{-3}$, from top to bottom.}
	\label{fig:m2m4}
\end{figure}

Introducing the operator $\op{D}_m = (r\frac{\partial }{\partial r}+m)$ allows us to write the laplacian as 
$\nabla^2 = r^{-2}\op{D}_{-m}\op{D}_{m}$.  We may then define the scalar function $C^{(m)} = \op{D}_m A^{(m)}$, and express the azimuthal terms of \eq\ (\ref{eq:u1}) as 
\be
   \label{eq:u}{\partial\over\partial t}{U}^{(m)} = -\oma^2\op{D}_{-m} C^{(m)}
 \ee
Operating with $\op{D}_m$ on \eq\ (\ref{eq:a}), gives
 \be
   \Frac{\partial C^{(m)}}{\partial t} = - \op{D}_m U^{(m)} + \eta \op{D}_m\Bigl(\Frac{1}{r^2}\op{D}_{-m}C^{(m)}\Bigr)
 \ee
 Changing variables to $s=\ln(r/\Delta)$, and letting $f^\prime$ denote $\partial f/\partial s$, we arrive at these telegrapher-like equations:
\begin{eqnarray}
   \label{eq:cmd}
   {\partial\over\partial t} C^{(m)} & =& -U^{\prime(m)} - m U^{(m)} + {\eta\over\Delta^2} 
   e^{-2 s}\Bigr[ C^{\prime\prime(m)}-2 C^{\prime(m)} - (m^2 - 2 m)C^{(m)}\Bigl] \\
   \label{eq:umd}
   {\partial\over\partial t}U^{(m)} &=& -\oma^2(C^{\prime(m)} - m C^{(m)}) ~~.
 \end{eqnarray}
The uniform resistivity $\eta$ is turned on at $t=0$, disrupting the current sheet.  We numerically solve the equations of motion using a finite difference method: \eq\ (\ref{eq:umd}) is updated explicitly, while for \eq\ (\ref{eq:cmd}) we employ an operator splitting method, updating the first two terms explicitly and the diffusive term implicitly.  We have a stationary initial condition, where $U^{(m)}(s,0)=0$.  
 $C^{(m)}$, derived from the perturbation $A_1$, has $r<\Delta$ initial condition 
 \be
   C^{(m)}(s,0) = \Frac{16 I_{\rm cs}}{\pi}\sum_{n=1}^\infty \Frac{n!!}{(n^2-2n)(n-m)(n-1)!!}e^{ns} ~~,
 \ee
 while outside the current sheet
 \be
   C^{(m)}(s,0) =\left\{ \begin{array}{lcl}-I_{\rm cs} &~~,~~&m=0\\0 &~~,~~&m\geq 2 ~~,\end{array} \right.
 \ee
 since $\op{D}_mr^{-m}=0$, for $m\ge2$.  
This regime well reproduces Fig 3.~\ from \citet{Longcope2007e} for the $m=0$ mode.  The $m\ge 2$ modes, however, are rapidly driven towards zero, even when $\eta = \oma \Delta^2$, as seen for $m=2,4$ in \fig\ \ref{fig:m2m4}.  The modes do propagate outwards for longer times, though not at a discernible level.  Larger $m$ modes are driven towards zero progressively faster.  Magnetic energy contributed by the terms $m\ge2$ is therefore dissipated locally by the resistivity, while the axisymmetric contribution leads to FMWs.

\end{document}